\documentclass{article}

\usepackage[english]{babel}
\usepackage[letterpaper,top=2cm,bottom=2cm,left=3cm,right=3cm,marginparwidth=1.75cm]{geometry}

\usepackage[dvipsnames]{xcolor}

\usepackage{amsmath,bbold,amssymb}
\usepackage{graphicx}
\usepackage[colorlinks=true , citecolor=ForestGreen, linkcolor=blue]{hyperref}

\renewcommand\l{\lambda}
\newcommand\w{\omega}

\title{\vspace*{-2ex}
Can we make sense out of negative kinetic energy?}

\author
{Alessio Maiezza$^{1,2\ast}$, Fabrizio Nesti$^{1,2\dagger}$\\[2ex]
\normalsize{$^{1}$ Dipartimento di Scienze Fisiche e Chimiche, Universit\`a} \\
\normalsize{degli Studi dell'Aquila, via Vetoio, I-67100, L'Aquila, Italy,}\\[.5ex]
\normalsize{$^{2}$ INFN, Laboratori Nazionali del Gran Sasso, 67010 Assergi, L'Aquila, Italy,}\\[.5ex]
\small{E-mail: $^{\ast}$alessiomaiezza@gmail.com, $^{\dagger}$fabrizio.nesti@univaq.it}\\[1ex]
}

\begin{document}
\maketitle

\begin{abstract}
\noindent Many Quantum Field Theory models, either designed to improve the ultraviolet behavior via higher derivative terms or aimed at providing unified descriptions of Gravity, inevitably introduce negative-energy 'ghosts' states. These states raise fundamental questions regarding the stability of the theory. In this study, we investigate the time evolution of simplified quantum mechanical models consisting of two coupled normal and ghost subsystems. Although the full spectrum of these models is unbounded from below and, therefore, lacks a true ground state, we focus the time evolution of a 'would-be ground state': a zero-energy eigenstate of the decoupled system. By analyzing its metastability under interactions, we gain insights into the dynamics of ghost-like states. Contrary to the standard paradigm, our results demonstrate that, across a broad range of couplings, the would-be ground state does not decay and remains dynamically stable over long times. This provides evidence towards resolving the stability problems of Hamiltonians that are unbounded from below. We also discuss the connections with the corresponding classical and semiclassical dynamics.
\end{abstract}

\tableofcontents

\section{Introduction}

The most fundamental description of reality so far is based on quantum field theory (QFT), with the Standard Model of particle physics standing as a specific, renormalizable realization. This successfully accounts for the strong and electroweak interactions, setting aside gravity. Such a limitation, among others, has suggested that not only the Standard Model, but also QFT in its conventional form may be incomplete and that a more comprehensive theoretical structure could be needed. One possible route toward such an extension is to improve the ultraviolet (UV) behavior of the theory by introducing higher-derivative operators.
While these terms enhance the ultraviolet behavior of the theory by rendering propagators more convergent, they generically lead to an Ostrogradsky instability manifested by a Hamiltonian unbounded from below and the emergence of ghost degrees of freedom (see e.g.,~\cite{Smilga_2005,PAV_I__2013,Kaparulin_2014,Motohashi_2015}). A consistent treatment of these ghost states would significantly extend the power of QFT to describe all fundamental interactions, including gravity~\cite{Salvio_2014,Salvio_2018}.

Ghost degrees of freedom can also arise in contexts unrelated to higher derivatives. Already at the classical level --  before addressing the challenges of quantization --  unified models of gravity with electroweak interactions (Graviweak) or with all forces (GraviGUT)~\cite{Cahill:1978ps,Cahill:1982zf,Percacci:1984ai,Percacci:1990wy,Maraner:2003sq,Nesti:2007ka,Nesti:2007jz,Nesti:2009kk,Maiezza:2021dui} lead to degrees of freedom with negative kinetic energy. Here, the nature of the ghosts differs from those arising from higher derivatives, since the ghost degrees of freedom arise already at the group-theoretical level, namely, in writing a unified Lagrangian for non compact unifying gauge groups. Other approaches try to exploit the ghost nature in order to stabilize the radiative corrections to the cosmological constant~\cite{Kaplan:2005rr,Craig:2024rdg}.

All these frameworks raise the fundamental question of whether ghost states make sense physically. In general, negative kinetic energy leads to a Hamiltonian which is unbounded from below, which in turn implies a possible instability. Therefore, technically, the problem is primarily formalized in terms of stability. In addition, the concept of instability at the classical and quantum level are intrinsically different, where the classical instability (runaway or blow-up) in this arena has been studied to some extent (see e.g.\ a recent review in~\cite{Errasti_D_ez_2025}). 
At the quantum level the situation is still uncertain, and a full answer  is extraordinarily challenging.
Still, useful insights can be gained by studying simplified toy models, which is the approach of the present work. 

We focus here on Quantum Mechanics (QM) and adopt a toy model of a coupled ghost system, represented by two oscillators, where one appears with the wrong sign in the Hamiltonian, coupled by a potential term,
\begin{equation} \label{eq:system}
H= \frac12 \Big[p_x^2-\omega_1^2 x^2\Big]-\frac12 \Big[p_y^2-\omega_2^2 y^2\Big]
+V(x,y)  \,.
\end{equation}
It is well-known that this system is closely related to the Pais–Uhlenbeck Oscillator~\cite{PhysRev.79.145}, one of the simplest prototypes of a higher derivative model (see~\cite{Ilhan:2013xe} for a recent reconsideration). 

Since the Hamiltonian is not bounded from below, a generic interaction potential $V(x,y)$ is commonly expected to induce unstable solutions in which both oscillators can exchange energy without bound.  In the quantum analysis, one expects localized states to decay. The aim is thus to test this expectation, by analyzing the system at the quantum mechanical level.  

Previous studies of the quantum Pais-Uhlenbeck oscillators tended to resort to complexification via a Wick rotation for the ghost oscillator~\cite{Smilga:2008pr}, or getting rid either of its negative norm or energy states through modifications of the inner product\cite{Bender:2007wu,Salvio:2015gsi,Strumia:2017dvt,Holdom:2024onr}. Some examples of  ghost models with no runaway have been addressed and analyzed at classical~\cite{Deffayet:2021nnt,Deffayet:2025lnj} and also at quantum level, e.g.\ in~\cite{Deffayet:2026cnu}, although these possess additional integrals of motion and can be classified as integrable problems. A number of investigations were devoted also to non integrable systems~\cite{Smilga:2008pr,Smilga_2017}, although the issue of quantum stability is still open.

In this work, we consider non-integrable systems and without modifying either the Hilbert space structure or the Hamiltonian. By directly solving the Schrodinger equation, we investigate both the spectral properties and the time evolution of the states, and compare the resulting quantum dynamics with the corresponding classical behavior. In this way, we shall provide a partial answer to the long-standing question of whether the unboundedness of the Hamiltonian necessarily translates into a quantum dynamical instability.

\paragraph{The problem.}

We consider a quantum system consisting of two harmonic oscillators with unit masses and frequencies $\omega_1$ and $\omega_2$. For simplicity, we shift their Hamiltonians so that the quantum ground state energies vanish. The full system, including the coupling potential $V$, is described by the Hamiltonian ($\hbar=1$)
\begin{equation}
    \begin{split}
&\hspace*{10em}H= H_1-H_2 + V \,,\\[1ex]
&H_1 = -\frac12\frac{\partial^2}{\partial x^2} + \frac{\omega_1^2}{2} x^2 -\frac{\omega_1}{2}\,,\qquad
 H_2 = -\frac12 \frac{\partial^2}{\partial y^2} + \frac{\omega_2^2}{2} y^2 -\frac{\omega_2}{2}\,,
 %V&= \frac{\lambda}{2}\, x^2 y^2  \qquad\text{or}\qquad \frac{\lambda}{2} x^2 y^2\,{\rm e}^{-(x^2+y^2)}\,, \nonumber \\[1.3ex]
  \label{mainH}
\end{split}
\end{equation}%
and the time-independent Schroedinger equation reads
$H\, \psi(x,y) = E   \,  \psi(x,y)$. For our analysis, we will consider two cases, one of a "small" and one of a "large" interaction potential: 
\begin{equation}
 V_{\text{small}}(x,y)=\frac{\lambda}{2} x^2 y^2\,{\rm e}^{-(x^2+y^2)}\,,
 \qquad    
 V_{\text{large}}(x,y)=\frac{\lambda}{2} x^2 y^2\,.
\label{eq:VsmallVlarge}
\end{equation}
The first interaction decays exponentially for large coordinate values and, therefore, leaves unchanged the asymptotic structure of the Scrodinger equation. Thus one may expect an analytic dependence of the spectrum and the eigenstates on the coupling $\lambda$. The second interaction alters the asymptotic structure of the Hamiltonian and is thus a genuinely non-perturbative modification of the system. 
We shall analyze first  the "small" case, and then  the more challenging "large" one, which was the subject of previous attempts~\cite{Gross_2021}.

In the decoupled case, $V=0$, it is easy to identify the zero energy state, which coincides with the ground state of both oscillators. Although the Hamiltonian is unbounded from below and hence has no true ground state, we shall refer to this zero-energy state as the \emph{would-be} ground state, $|\bar v_0\rangle$. The reason is that this is the closest resemblance of the zero-point-energy state or, in a QFT analogy, the vacuum, where both oscillators are prepared in their lowest energy, or least excited, configurations.  

The would-be ground state $|\bar v_0\rangle$ is an eigenstate of the decoupled Hamiltonian, but it starts to evolve when the interaction  $V$ is turned on. We  now  pose a well-defined question of what happens to this state under time evolution.  In particular, we aim to assess whether the would-be ground state is unstable/metastable; namely, whether its survival probability vanishes in time.
Using the eigenvectors and eigenvalues of the coupled system, $|v_j\rangle$, $E_j$, any initial state $| v\rangle$ can be expanded as $|v\rangle=c_j|v_j\rangle$, with overlap coefficients  $c_j=\langle v_j|v\rangle$. Then its survival probability in time is
\begin{equation}
\label{eq:P}
P_{v} (t)= \big|\langle v|v_t\rangle\big|^2 = \bigg|\sum_j \left|c_j\right|^2 e^{-i E_j t}  \bigg|^2  \,,
\end{equation}
assuming for the moment a discrete spectral decomposition.

Since stability or meta-stability of the vacuum is a fundamental problem, this article is mainly dedicated to estimating the survival probability for the would-be ground state, $|\bar v_0\rangle$. However, we shall briefly compare also with the fate of excited states. Therefore, we will try to provide an answer to the stability question and discuss the relative technical difficulties.

We anticipate that the answer is partially positive, namely, the would-be ground state does not decay, at least in  some range of the coupling $\lambda$. Since our analysis relies on numerical methods, the system evolution and stability can only be tested over finite time intervals. Strictly speaking, our results establish long-lived \emph{metastability} rather than an absolute stability. Interestingly, we note certain classical-to-quantum correspondences pointing to a smooth semiclassical transition.

\section{Decoupled quantum ghost system}\label{sub_ghost_QM_dec}

For $\lambda=0$, the harmonic oscillators are  decoupled: the system is separable and the eigenstates live in the product of spaces. The system is equivalent to the free Pais–Uhlenbeck Oscillator~\cite{Smilga:2008pr}.
Requiring normalizable eigenstates, each oscillator has the standard set of countable eigenstates. 
However, because the second oscillator enters the Hamiltonian with an overall minus sign, the total Hamiltonian is not positive-definite. The spectrum in the decoupled case is
\begin{equation}\label{free_spectrum}
(\lambda=0) \quad: \qquad  \bar E_{n_1,n_2}= n_1 \omega_1 - n_2 \omega_2        \qquad   \bar \psi_{n_1,n_2}(x,y)= \psi_{n_1}(x) \, \psi_{n_2}(y) \,,
\end{equation}
with $n_{1,2}$ positive integers. Thus, $\bar E$ ranges from $-\infty$ to $+\infty$.

Formally, the spectrum in \eqref{free_spectrum} is qualitatively different depending on whether $\omega_1$, $\omega_2$ are commensurable or not. In the case they  are commensurable, the spectrum is discrete, and each eigenvalue is infinitely degenerate. In the case $\omega_{1,2}$ are incommensurable, the spectrum is dense in $\mathbb R$. We will study the latter case by adopting the simple choice
\begin{equation}\label{free_spectrum_2}
\omega_1=1,   \hspace{2em}   \omega_2=\sqrt{2},  \hspace{1em} \Rightarrow \hspace{1em}   \bar E_{n_1,n_2}= n_1  - \sqrt{2}\,n_2  \,.
\end{equation}
Although the values of $\bar E$ do not cover all $\mathbb R$, being just combinations of an integer and a multiple of $\sqrt{2}$,  any real value can be approximated arbitrarily well with suitable large values of $n_{1,2}$. This is guaranteed by Kronecker's approximation theorem~\cite{kronecker1884naherungsweise}. For instance, the energy can be made arbitrarily close to $0$ by choosing a progression of couples $n_{1,2}$ that approximate  $\sqrt{2}$, like the continued fractions satisfying $n_1/n_2 \sim \sqrt{2}$. 

As detailed below, the spectrum becomes dense in $\mathbb R$ (not  continuous) as $n_1$ and $n_2$ range up to infinity (no UV cutoff). Therefore, one has states with arbitrarily small total energy, with $n_{1,2}$ both large. In general, the properties of the spectrum around (any) low energy depend on the UV regime of the model. This is traced back to the minus sign of the ghost oscillator in~\eqref{mainH}, which causes an ultraviolet-infrared mixing due to the cancellations between $H_1$ and $H_2$. This conflicts with the decoupling paradigm asserting that very separate scales do not participate together in physical phenomena. Not surprisingly, departing from standard systems, effects of IR-UV mixing may signal conceptual and technical troubles.

\subsection{Density of states.}

It is worth discussing the density of states in more detail. As said, the Kronecker's approximation theorem ensures  that $n_1-r\, n_2$ can approximate  any real number if $r$ is irrational.  For instance in our case with $r=\sqrt{2}$, for any real energy $E$ and any given  small positive $\Delta E$, there exist $n_1$, $n_2$ such that $|n_1-\sqrt{2} n_2  -  E | < \Delta E/2$.

Generically, one has
\begin{equation}
\left\vert n_1 - r \,n_2 -\frac{E}{\omega_1} \right\vert \le \frac{\Delta E}{2\,\omega_1} \,,
\end{equation}
with $r\equiv \omega_2/\omega_1$ irrational, i.e.,\ $\w_1$, $\w_2$  non commensurable.

To evaluate the density of states, we have to consider the number of states satisfying the above condition. Considering the first quadrant of the \emph{real} $n_1$, $n_2$ plane,  the above condition defines a diagonal strip. The Kronecker approximation theorem then implies that for any nonzero $\Delta E$ there are infinitely many discrete states falling inside this strip. Therefore, the density of states requires an UV regularization to be defined. 

Let us introduce an UV cutoff by defining, for simplicity, a square domain extending up to maximal energy $\Lambda$ for both energies,
\begin{equation}
    \label{eq:cutoff}
    E_{1,2}=\w_{1,2}n_{1,2}\in [0,\Lambda]\,.
\end{equation}
Then the lattice has surface density = $1/\w_1\w_2$ and is limited to $n_{1,2\,\text{max}} <\Lambda/\omega_{1,2}$. The number of states falling in the strip is proportional to its area, and the density of states to its length. Near $E\approx 0$, one finds 
\begin{equation}
\rho_\Lambda(E\approx 0) = \frac{\Lambda}{\omega_1\omega_2} \,.
\label{eq:rho_geom}
\end{equation}
For generic $E$, the density of states in the presence of the rectangular UV cutoff is precisely defined as
\begin{equation}
\rho_\Lambda(E) = \sum_{(n_1, n_2) \in \mathcal{D}_{\text{UV}}} \!\!\delta\left(E_1-E_2-E\right) \,,
\end{equation}
where $\mathcal{D}_{\text{UV}}$ is the lattice inside the square domain  $0<E_{1,2}<\Lambda$. 
By passing to the dimensionless variables $\xi_i=E_i/\Lambda$ and $\xi=E/\Lambda$, we have 
\begin{equation}
\rho_\Lambda(E) = \frac\Lambda{\w_1\w_2}\sum_{(\xi_1, \xi_2) \in \mathcal{D}_{\text{UV}}} \!\!\delta\left(\xi_1-\xi_2-\xi\right) \Delta \xi_1 \,\Delta \xi_2\,,
\end{equation}
where the domain of the $\xi_i$ variables covers the unit square and has shrinking spacings $\Delta \xi_i = \w_i/\Lambda$. Thus, for large $\Lambda$, the above appears as a Riemann sum and one is led to pass to the continuum:
\begin{equation}
\rho_\Lambda(E) \simeq \frac\Lambda{\w_1\w_2}  \int_0^1 \!{\rm d}\xi_1 {\rm d}\xi_2  \,\delta(\xi_1-\xi_2-\xi) = \frac\Lambda{\w_1\w_2} \left( 1-|\xi|\right)= \frac{\Lambda -|E|}{\w_1\w_2} \,.
\label{eq:rhoofE}
\end{equation}
For energies $E \ll \Lambda$, one exactly recovers expression \eqref{eq:rho_geom}.

\medskip

In the UV limit the spectral density \eqref{eq:rhoofE} grows with $\Lambda$, showing the trend towards a dense spectrum. In addition, $\rho_\Lambda(E)$ divided by $\Lambda$ becomes constant in $E$, i.e.,\ the spectrum appears uniform in $E$, as if one zooms into small $|E|$. 
This suggests that one may consider the limit of large $\Lambda$ as an UV analogue of the IR case of a particle in the 2D box. Considering a 2D box of size $L$, the density of states  $L^2\, m/2\pi$ is constant in energy and proportional to the growing IR cutoff volume $L^2$. When the spectrum is taken to approximate a continuous spectrum, the relevant intensive physical quantity is the density of states per unit volume, $m/2\pi$.
It is thus worth discussing further the partial analogy with this case. 
We can try to renormalize the density of states $\rho_\Lambda(E)$ by isolating the cutoff-independent part, namely, dividing by the effective "volume" of contributing states in energy space, namely $\Lambda$:
\begin{equation}
g(E) \equiv \lim_{\Lambda \to \infty} \frac{\rho_\Lambda(E)}{\Lambda} = \lim_{\Lambda \to \infty} \frac{1 - \frac{|E|}{\Lambda}}{\omega_1\omega_2} = \frac{1}{\omega_1\omega_2}  \,.
\label{eq:gofE}
\end{equation}
Therefore, like for the free particle in a 2D box, the total state counting diverges with the cutoff ($\Lambda$ for the ghost, $L^2$ for the box), but the density per unit cutoff yields a dense---not yet continuous---and constant spectrum.  

The crucial difference between the two system is in the limiting procedure, concerning the states. The standard rescaling of states, in order to pass from the discrete orthogonality $\langle E \vert E' \rangle = \delta_{E,E'}$ to the truly continuous one, requires one to use the density of states: 
\begin{equation}
    \vert\tilde{E}\rangle\equiv \sqrt{\rho_\Lambda(E)}|E\rangle=\sqrt{\frac{\Lambda-|E|}{\omega_1\omega_2}} \, \vert E \rangle\quad \implies \quad \langle \tilde E \vert \tilde E' \rangle = \delta(E-E')  \,,
\end{equation}
where the delta is now the Dirac delta. For small energy $|E|\ll\Lambda$, this amounts to a constant rescaling, like it happens for the spatial box. However, the  difference is that, while in the 2D box the limit $L \to \infty$ leads to well defined eigenstates (plane waves, possibly defined rigorously via generalized states in a rigged Hilbert space), in our system the limit cannot be taken directly on the states $\vert\tilde{E}\rangle=\sqrt{\frac{\Lambda}{\omega_1\omega_2}} \, \vert E \rangle$, because their wavefunctions diverge everywhere for $\Lambda\to \infty$. This is obviously the signal that the ghost system  possesses a dense \emph{point} spectrum that \emph{can not} be transformed into an absolutely continuous one as the free particle.

We have thus a quite peculiar quantum system, in which an UV cutoff is essential in defining the properties of the spectrum, even at much lower energies. This lack of decoupling between the UV and IR physics is the root of a number of consequences, which persist in the coupled case. For this purpose, we will later numerically confirm the spectral density for the decoupled system, as a setup for the more interesting coupled system, which we can only probe through numerical approximations.

\section{Coupled quantum ghost system}

Physics becomes even more interesting when an interaction between the two subsystems $\l$ is turned on. As stated in the introduction, our main question is whether the presence of this coupling can lead to blow-up or runaway behaviors, which, at the quantum level, correspond to the decay of an initial state into a combination of other eigenstates, eventually with zero survival of the initial configuration.  In particular, one expects here an infinite number of possible states each having large and opposite energies in the normal and ghost oscillators. 

Therefore, we focus on the setup where one prepares an initial state and checks its survival in time. As mentioned in the introduction, we adopt as initial state the \emph{unperturbed} ground state of both oscillators, i.e., the state relative to $\bar E=0$, which has $(n_1,n_2)=(0,0)$. In the interacting system, this initial state is only an approximate eigenstate of the full Hamiltonian, and we can check its survival probability in time by studying its time evolution. 

\medskip

As a guiding example of a decaying state, one can mention the textbook case of the metastable trap, with a finite height barrier: one can prepare an approximate eigenstate solely in the inner region, exact only in the limit of infinitely high barrier. With a finite barrier, this state decays to infinity due to some probability current tunneling in the outer region. One also knows that this decay can only happen because the inner region is coupled via the finite barrier with the outer one, which has a \emph{continuous} spectrum. Due to this coupling, even the full spectrum is continuous: the discrete inner states become metastable with some small width $\Gamma$ of the  density of states in energy ($\Gamma$ being determined semiclassically by the barrier properties). The initial state leaks to the outside with a lifetime $1/\Gamma$. When checking the lifetime, the decay is possible due to an infinite number of nonzero overlap integrals with the full continuous eigenstates. For instance, this case  was reviewed in detail in~\cite{Andreassen:2016cvx}, where it is shown that various regimes appear: an initial leaking, then an exponential decay, followed by a final power-like decay to zero (see also~\cite{Ram_rez_Jim_nez_2019}).  On the other hand, when also a further external wall is added, the initial state does not decay to zero but  there is non-null probability current returning to the false vacuum, leading to oscillations instead of decay. Technically, this can be seen to be consequence of the  discrete spectrum in the outer region.

Therefore, the fact that our spectrum is effectively dense opens the possibility that the survival of our $(0,0)$ state will decay to zero, due to the leaking into an infinite superposition of the true eigenstates. This would reproduce the expected runaway into positive plus negative energies.  We will see that our results do not support this conclusion, but rather show an oscillating behavior. 

\medskip

All the physics is ultimately encoded in the eigenvalues  and eigenvectors of the coupled system, which we name $E_j$ and $|v_j\rangle$, as opposed to the free system for which we reserved the barred notation, $|\bar v_j\rangle$, $\bar E_j$ as in \eqref{free_spectrum_2}. Here the index $j$ will be different in each case, just labeling the (countable) spectrum under consideration.  
\medskip

In the coupled case and for either the "small" or "large" interactions in \eqref{eq:VsmallVlarge}, there is no analytical knowledge of the spectrum as in the decoupled case. To attack the problem, there are a few possible alternatives. One may resort to truncated perturbation theory, or one may approach the problem numerically. Both  methods have intrinsic limitations that we will discuss, and we shall follow both paths in the attempt to capture different aspects of the problem. 

Before unfolding these methods, 
we  wish to  pinpoint  a couple of conceptual peculiarities of the problem under consideration:

\begin{enumerate}

    \item[(A)] A first comment regards a possible perturbation expansions in $\lambda$. The problem at hand falls within the theory of linear operators, developed in \cite{Kato:1966:PTL}. That seminal book presents two main types of perturbation approaches. The first, \emph{analytic perturbation theory} (see Chapter~7), is quite restrictive: it relies on the concept of \emph{relative boundedness} of the perturbing operator, and guarantees convergence only for isolated eigenvalues. Although the relative boundedness condition is technically satisfied by our ``small'' (bounded) potential, the dense nature of the unperturbed spectrum prevents the rigorous application of this framework. The ``large'' potential falls even further outside the realm of applicability, as it violates the relative boundedness condition itself due to its arbitrary growth for large $x$ and $y$. Note that the latter violation is not due to the negative-sign (ghost) contribution from $H_2$ in \eqref{mainH}; the analytic approach would be unsuitable even in the standard positive-sign case for such unbounded interactions.

    The second approach, covered in Chapter~8, involves \emph{asymptotic expansions}. The underlying assumption here is that there are isolated eigenvalues of the unperturbed operator. While this theory guarantees an asymptotic expansion for the perturbed eigenvalues in the standard (non-ghost) case, it does not strictly match our scenario since the unperturbed spectrum is dense. Notice that this makes our problem genuinely different from standard problems with unbounded potentials: in \eqref{mainH}, the cancellation between the kinetic energies of $H_{1,2}$ is responsible for the dense spectrum.

    Despite the absence of a rigorous theoretical foundation for perturbation theory in our context, we will attempt formal expansions in $\lambda$, aiming to extract partial insights into the system's behavior, merging and comparing with non-perturbative information obtained via numerical approaches.

    \item[(B)] The computational and physical necessity to introduce  a UV cutoff implies that one has a sequence of spectra (eigenvalues and eigenvectors) for the various choices of increasing cutoff.
    We will use the subscript $^{(n)}$ to label the approximations. The cutoff applies also to the decoupled system, so for instance we will denote as $v_j^{(n)}$ and $\bar v_j^{(n)}$ the eigenvectors in the coupled and decoupled systems, in the presence of a cutoff labeled by $n$. 

    The sequences ideally should approximate the 'true' values in the continuum limit.  One would thus  like to check the convergence of these approximations.   
    For instance a form of convergence may be formalized in terms of \emph{weak convergence} in a Hilbert space, $v_i^{(n)} \rightharpoonup v_i $.
    However, one faces a difficulty in checking the convergence of the sequence: due to the progressively dense spectrum as the cutoff grows, one can not identify the  states with the same $i$ between different values of $n$.  While this would be possible for standard systems, our case has the peculiar property that the eigenstates and eigenvalues get continuously reshuffled, as the spectrum becomes increasingly dense around any energy.
    
    In our system,  we are interested in the overlap coefficients between the coupled and decoupled states.    In particular we focus on overlap coefficients with the would-be ground state, i.e.\  we will test  for the existence of the limit of 
    \begin{equation}\label{weakH}
    \lim_{n\rightarrow \infty} \langle v_i^{(n)} \,|\, \bar v_0^{(n)} \rangle \stackrel{?}{=} \langle v_i \,|\, \bar v_0 \rangle \,.
    \end{equation}
    Also checking \eqref{weakH} is difficult because one can not track the states of a given $i$ for varying $n$. 
    We approach this problem of identifying states by choosing some particular ordering, namely by sorting, for each $n$, the approximate eigenstates $v_i^{(n)}$ by their overlap with the free state $\bar v_0^{(n)}$.  Then, also in the case of the large potential, we find that we can answer positively to \eqref{weakH} when $\lambda$ is sufficiently small, $\lambda \lesssim 0.05 $.
    For larger $\lambda$, an increasing number of transient eigenvalue crossings prevents the limit to be proven.

\end{enumerate}

Despite these issues, we anticipate that, for either choice of potential, we will find evidence that the spectrum remains dense  and that the survival probability displays a convergence for large $n$. The next section performs this analysis for the case of the small interaction potential, followed then by the case of the large potential. 

\subsection{Coupled ghost system: Small potential case}

In this first case, the potential has the form
\begin{equation}
    V=\frac\lambda2 x^2 y^2 {\rm e}^{-(x^2+y^2)}\,.
    \label{eq:potsmall}
\end{equation}
Since $V$ vanishes rapidly at infinity, it does not modify the asymptotics of the eigenvalue equation, suggesting that perturbation theory should give reliable results. We wish to test this expectation, and to investigate the time survival of the (0,0) initial state.
Let us first discuss the perturbative approach and later perform the full numerical analysis. 

\subsubsection{Perturbative approach}

The eigenvalues are expanded perturbatively in the coupling $\lambda$, and by using the bi-index $(n_1,n_2)$ of the free spectrum (\ref{free_spectrum}) one has
\begin{equation}\label{genpert}
E_{(n_1,n_2)} =  \bar E_{(n_1,n_2)} +\sum_{p} f_{p}(n_1,n_2)\lambda^p  \,,
\end{equation}
where $\bar E_{(n_1,n_2)}=\w_1 n_1-\w_2n_2=n_1-\sqrt{2} n_2$ are  the free eigenvalues.

The coefficients $f_p$ can clearly be calculated by standard perturbation theory, involving sums on intermediate free states and matrix elements of the interaction potential.  These may be approached analytically by recalling that the generating function of the free oscillator eigenfunctions (i.e. Hermite polynomials) is an exponential in $x$ and thus can be integrated with the gaussian potential \eqref{eq:potsmall}, separately for each coordinate.  Hence the full matrix elements of $V$ are just the tensor product of the two sectors,
\begin{equation}
V_{(m_1,m_2),(n_1,n_2)}\equiv \langle m_1,m_2|V|n_1,n_2\rangle=\frac\lambda2\,V^1_{m_1,n_1} V^2_{m_2,n_2}\,.
\end{equation}
In each sector  one  finds the generating function  for the matrix elements:
\begin{equation}
    V^{i=1,2}_{mn}=\langle m|x^2 {\rm e}^{-x^2}|n\rangle = \left.\frac{\partial^n_k\partial^n_q \,G_{\omega_i}(k,q)}{\sqrt{n! \,2^n}\sqrt{m! \,2^m}}\right|_{k=q=0}\,,\end{equation}
    where
    \begin{equation}
G_{\omega_i}(k,q)=\sqrt{\omega_i}\frac{1+\omega_i+2\omega_i(q+k)^2}{(1+\omega_i)^{5/2}}\, {\rm e}^{-\frac{k^2+q^2-2\omega_i kq}{1+\omega_i}} \,.
\end{equation}
We notice now that since the potential is not polynomial, even at finite perturbative order there is an infinite number of nonzero matrix elements entering the  sums on intermediate states. One can check that these matrix elements are not even suppressed for large $m$,$n$.  Therefore, one has to regulate the sums by truncating the Hilbert space in the UV, and check for convergence of the results when removing this regulator. Higher perturbative orders require increasingly larger regulator,  hitting the computational capabilities. For instance, when truncating each unperturbed oscillator spectrum to $n$, one has $\sim (4n^4)^{p-1}$ terms. This implies that,  e.g.,\ at fourth order and with $n=10$, one has $6\times 10^{13}$ terms. Remarkably, we are able to treat confidently the series up to the sixth order, while verifying independence from the cutoff.

For the $(0,0)$ state, after defining the first order shift as $\tilde\lambda =\lambda/16\sqrt{10+7\sqrt{2}}\simeq \l/71$,  we find
\begin{equation}
\label{eq:E0pertsmall}
   E_{(0,0)}= \tilde\lambda-0.0434241 \,\tilde\lambda^2-0.289552 \,\tilde\lambda^3
   -0.258875 \,\tilde\lambda^4
   -0.33841 \,\tilde\lambda^5
   -0.1020 \,\tilde\lambda^6+O(\tilde\l^7)\,.
\end{equation}
This truncated series will be useful to compare with the nonperturbative results below.  The series coefficients appear to be limited, suggesting a nonzero radius of convergence, namely, $\tilde\lambda\lesssim O(1)$. This is expected because the potential vanishes for large values of $x$ and $y$. Similar results appear when analyzing the diagonalizing matrix, where we find that unitarity is reliable up to  $\tilde\l\lesssim 0.5$, i.e.\ $\l\lesssim 30$.

On the other hand, the perturbative analysis can not answer the question whether the spectrum remains dense like the free one or not.  In  fact, in \eqref{genpert}, one would realize a dense spectrum of the $E_{(n_1,n_2)}$ for  growing $n_{1,2}$, by canceling the free $n_1-\sqrt{2} n_2$  with the large corrections $\l^p f_{p,i}(n_1,n_2)$. This breaks the expansion, i.e.,  the spectral density is genuinely nonperturbative.

\subsubsection{Numerical approach: Discretization}\label{subsec:discr:small}

Let us discretize the Hamiltonian \eqref{mainH} on a lattice of step $a$ made of $2n$ points in each $x$ and $y$ directions. We represent $\psi(x,y)$ on the discretized lattice as $\psi_{ak,aj}$, with $k,j=0\ldots 2n$, and with null conditions at the boundaries. 

This  approach has two usual intrinsic regulators: a UV one, $\Lambda\sim 1/a$, and an IR one given by the volume size, $L=2na$, which we take equal in the two variables.
With the choice $m=\hbar=1$, $\omega_1=1$, and $\omega_2=\sqrt{2}$, the parameters $a$, $L$ are dimensionless.
We refine the lattice by raising $n$ and lowering $a$, scaling jointly the UV and IR cutoffs as
\begin{equation}\label{cutoff}
 a^{-1}=C\, n^{1/2} \qquad L\equiv 2\, n\, a    \sim n^{1/2} \,. 
\end{equation}
Thus, as $n$ grows, both the UV and the IR regulators are lifted as $n^{1/2}$.  
This specific direction in the $a$-$L$ plane is chosen, among others, after testing the convergence of the discretized spectrum in the free case, where one knows the exact spectrum in the continuum and infinite volume limit.  The power and the value $C=1.2$ in \eqref{cutoff} gives the best convergence.

The two-dimensional kinetic term is written by discretizing the second derivatives\footnote{For better convergence, we have used an improved finite-difference scheme of derivatives at $\mathcal{O}(a)^4$, namely:
\begin{equation}\label{eq:discrete_kin_2}
\Delta^2_x\, \psi_{ak,aj}  = \frac{\psi_{a (k-2),a j}-16 \psi_{a (k-1),a j}+30 \psi_{a k,a j}-16 \psi_{a (k+1),a j}+\psi_{a (k+2),a j}}{24
   a^2}\,. 
\end{equation}
}
\begin{equation}
\label{eq:discrete_kin}
H_{kin}\, \psi_{ak,aj} = - \frac12(\Delta^2_x-\Delta^2_y) \,\psi_{ak, aj} 
\end{equation}
and the potential parts give 
\begin{equation}\label{discrete_pot_small}
H_{pot}\, \psi_{ak,aj}  = \left[\frac{1}{2} \omega _1^2 \left(a^2 j^2-\frac{1}{\omega _1}\right) -\frac{1}{2} \omega _2^2 \left(a^2
   k^2-\frac{1}{\omega _2}\right) + \frac{\lambda}{2} a^4 j^2 k^2 e^{-a^2 (j^2 +k^2)} \right] \psi _{a k,a j} \,,
\end{equation}
where we have subtracted the zero point energies consistently with \eqref{mainH}. 

With this discretization, we compute the spectrum of the full system $H_{pot}+H_{kin}$ of \eqref{eq:discrete_kin} and  \eqref{discrete_pot_small}, namely, the full set of the eigenvalues $E^{(n)}_i$ and eigenvectors $|v^{(n)}_i\rangle$, for each chosen $n$. 
The Hamiltonian to be diagonalized has dimension
\begin{equation}\label{Dim}
(2 n-1)^2 \times (2 n-1)^2 \sim 16 n^4\,,
\end{equation}
thus the problem becomes quickly resource consuming. 

In parallel, in order to have our initial would-be ground state $|\bar v_0^{(n)}\rangle$ in the discretized system, we have to compute also the spectrum of the free system ($\lambda=0$).  This is the direct product of the two one-dimensional discretized subsystems, thus drastically simpler. In particular, for each $n$, our initial $|\bar v_0^{(n)}\rangle$ is the product of  the ground states of the two decoupled one-dimensional \emph{discretized} systems. Note that this  exactly identifies the would-be ground state, even though its energy is not exactly zero (in the discretized system) and other states may end up having even smaller absolute energy. 
As $n$ grows, the UV and IR cutoffs~\eqref{cutoff} are removed and the spectrum of the decoupled systems approaches  the free ones \eqref{free_spectrum_2}. Thus the energy of $|\bar v_0^{n}\rangle$ approaches zero (further details on the cutoff removal are presented below). 

After calculating both spectra, we can evaluate the overlap coefficients
\begin{equation}\label{CC}
c^{(n)}_i= \langle v^{(n)}_i|\bar v_0^{(n)}\rangle \,.
\end{equation}
We order the states in the full spectrum by decreasing $|c_i^{(n)}|$, so that $|v_0^{(n)}\rangle$ is chosen as the state with highest overlap $c_0^{(n)}$ with the initial state $|\bar v_0^{(n)}\rangle$.

\begin{figure}[t]
\centerline{\includegraphics[width=0.97\columnwidth]{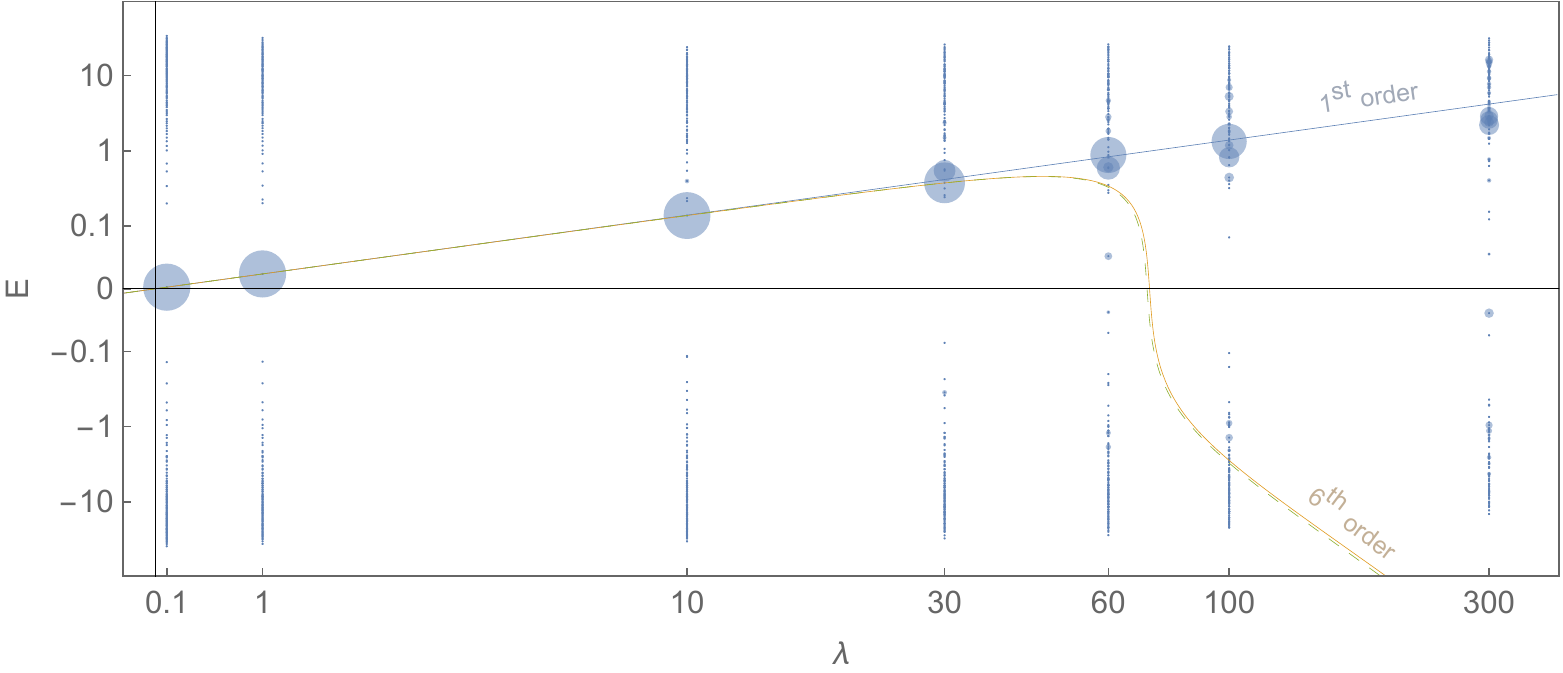}
}
\vspace*{-1.5ex}
\caption{Diagonalization for  $\lambda=0.1, 1, 10, 30, 60, 100, 300$ (and $n=50$).  For each $\l$, eigenvalues are shown  as blue dots, with the size proportional to the overlap coefficient $c_i^{(n)}$. The majority of eigenvalues, having negligible overlap, are still shown with minimal dot size.  
The blue(yellow) curves represent the  ground state energy in perturbation theory at $1^{\text{st}}$($6^{\text{th}}$) order. The plot can be extended to negative $\lambda$, with an analogous (non symmetric) result. Notice the ``sinh" scales on both axes.}
\label{fig:pallinismall}
\end{figure}

The numerical results for exact eigenvalues and overlap coefficients are shown in Figure~\ref{fig:pallinismall}. There, eigenvalues are also compared with the perturbative result~\eqref{eq:E0pertsmall}. The spectrum coincides exactly with the perturbative curve for small $\lambda\lesssim 30$ ($\tilde\lambda<0.5$). In this perturbative regime,  the state with maximal (circa unitary) overlap with the would-be ground state coincides with the perturbed ground state.

For larger $\lambda$, i.e.,\  $\lambda\gtrsim 30$, overlap with more states start to appear. 
Around this value, one also notices that the perturbation theory ground state departs from the most overlapping state, signaling that  we approach the strong coupling regime, beyond the radius of applicability of perturbation theory, as mentioned before. 
In the deep non-perturbative regime, for $\lambda> 50$, a curious phenomenon appears: the eigenvalues with largest overlap persist in a qualitative alignment with the simple first order perturbation.

%%%%%%%%%%%%%%%%%%%%%%%%%
 \begin{figure}[t]
 \centerline{
\includegraphics[width=.5\columnwidth]{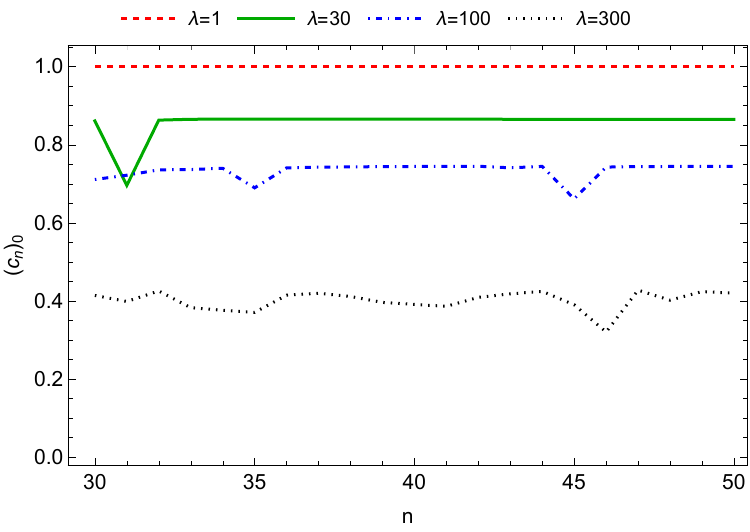}
}
\vspace*{-1.5ex}
\caption{Convergence of the largest overlap coefficient   with the initial state $c_0^{(n)}$, for increasing $n$.  A good convergence is obtained up to $\lambda=30$, whereas the overlaps reach a constant value. The convergence is slightly spoiled for larger value of the coupling.}
\label{plotweakHsmall}
\end{figure}
%%%%%%%%%%%%%%%%%%%%%%%%%%%

\medskip

An  aspect to be discussed is also the convergence with $n$ of the overlap coefficients introduced in point (B) above: here, we find that  we can  verify the convergence of $c_i^{(n)}$ only for small $\l$. 
In Figure~\ref{plotweakHsmall} we plot the $c_i^{(n)}$ for a sequence of values of $n$, and a few choices of $\l$. 
One can notice that for small couplings, $\lambda<30$ the coefficients converge, while for $\l\gtrsim 30$ they  oscillate slightly  around some constant value, still indicating a signal of convergence that may be improved by increasing $N$.

\subsubsection{Survival probability}

\begin{figure}[t]
 \centerline{
\includegraphics[width=.63\columnwidth]{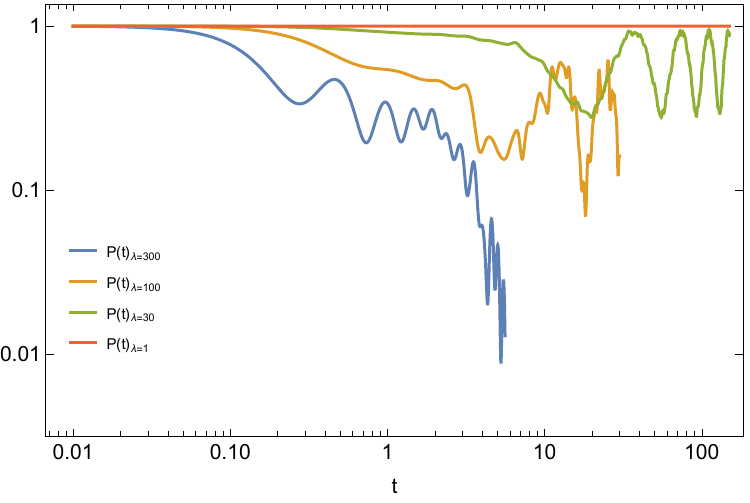}
}
\vspace*{-1.5ex}
\caption{Evolution in time of the survival probability. The plot is cut at $t=100$ but the curves for $\l=1$, 30  are reliably cutoff independent to much larger time. Instead the curves for $\l=100$, 300 are stopped at time until which cutoff independence is verified (see text). In every  case, instead of a decay in time, we uncover an oscillatory regime at late time, with residual probability which is increasingly suppressed for larger $\lambda$.}
\label{allsmall}
\end{figure}

We proceed to evaluate the survival probability~\eqref{eq:P} for given $n$,
\begin{equation}\label{discrete_P}
P_n(t)= \left|\sum_j \left|c^{(n)}_j\right|^2 e^{-i E^{(n)}_j t}  \right|^2  \,.
\end{equation}
The numerical results are illustrated in Figure~\ref{allsmall}, for $\lambda=\,1,\,30,\,100,\,300$, and for $n=50$. The survival probabilities are plotted only up to some maximal time, chosen in order to maintain a negligible dependence on the cutoff $n$ (we ask a deviation between different choices of $n$s ($n\geq35$) within $1\%$). An exception are the cases of small  $\lambda=1,\,30$, where the evaluation can be trusted for much larger time, and the axis is cut at $t=100$ for illustrative purposes only. Thus the results for $P(t)$  show good numerical convergences properties. 

%%%%%%%%%%%%%%%%%%%%%%%%%
%%%%%%%%%%%%%%%%%%%%%%%%%%%

The results for $P(t)$ in Figure~\ref{allsmall} show that the survival probability does not drop to zero.   For small coupling one notices a small reduction plus residual oscillations, and the result can be reliably trusted to very large times, at machine precision limits.
The same behavior is found for large coupling ($\lambda\gtrsim 100$), but the residual oscillating survival probability  drops to much smaller residual values and the reliability in time is much more limited. 

In summary, we wish to conclude that the considered ghost system does not decay, when the coupling $\lambda$ is  small or even moderate. Roughly, at least 50$\%$ of the initial would-be ground state survives at any time for $\lambda\leq 30$, and nearly the whole initial state remains for $\lambda=\mathcal{O}(1)$. It is worth to emphasize that, particularly when $\lambda=\mathcal{O}(1)$, the behavior is insensitive to the cutoff already at small $n$, supporting the robustness of the conclusion.

\subsubsection{Extracting energy?}

An interesting question concerning a system coupling a normal and a ghost subsystem is whether energy can be extracted from the latter. 
While the total energy is conserved,
\begin{equation}
E_{tot}=\langle \psi(t) \vert H \vert \psi(t) \rangle = \text{const}\,,
\end{equation}
and for our initial would-be ground state, one has $E_{tot}=\tilde\l$, the two subsystems clearly can exchange energy. 
Therefore, we  evaluate the mean energy of each subsystem as a function of time, 
\begin{align}
\label{eq:E1E2}
E_1(t)&= \langle \psi(t) \vert H_1 \vert \psi(t) \rangle \,, \nonumber \\
E_2(t)&= -\langle \psi(t) \vert H_2 \vert \psi(t) \rangle \,, 
\end{align}
where from here onward we stick to the convention $E_1>0$ and $E_2<0$.  

\begin{figure}[h]
 \centerline{
\includegraphics[width=.5\columnwidth]{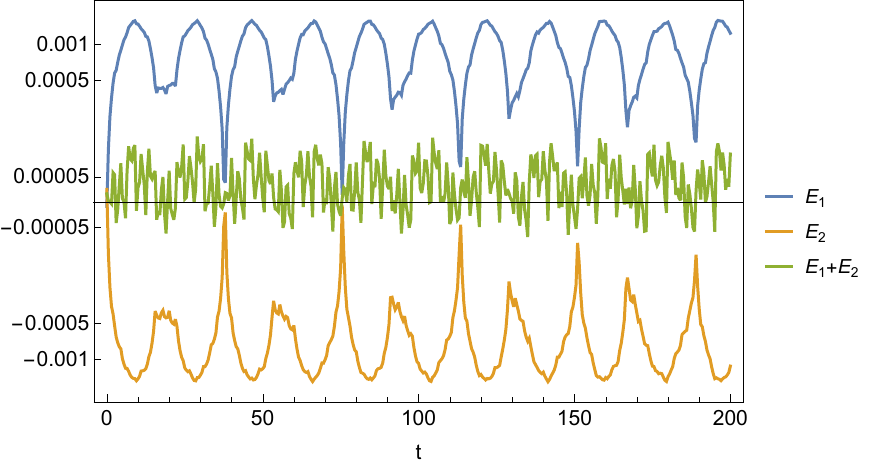}
%}
%\vspace*{-1.5ex}
%\caption{}
%\label{fig:energy_small_lam1}
%\end{figure}
%%%%%%%%%%%%%%%%%%%%%%%%%%%%
%
%
%%%%%%%%%%%%%%%%%%%%%%%%%%
% \begin{figure}[b]
% \centerline{
\includegraphics[width=.5\columnwidth]{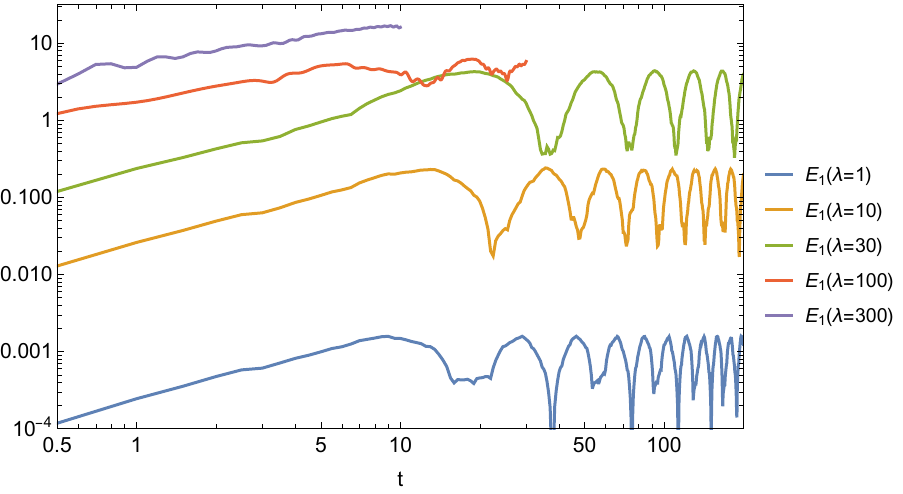}
}
\vspace*{-2ex}
\caption{Energy budget for the two oscillators in the small potential case. Left: separate and summed energy of the two oscillators in time, for $\l=1$. Right: energy gain of the first oscillator in time, for $\l=1$, 10, 30, 100, 300.}
\label{fig:energy_small}%_allam}
\end{figure}

From the discretized approach and for $\lambda=1$, we obtain the results shown in the left frame of Figure~\ref{fig:energy_small}. We observe that the standard (positive energy) subsystem does not steadily extract energy from the ghost; rather $E_1(t)$ shows an oscillating behavior. The same happens for $E_2(t)$, which is negative.  An oscillating behavior holds also for their sum $E_1(t)+E_2(t)$, where one can observe the cancellation, which is not exact.  The latter in fact also provides information about the interaction energy $E_{int}(t)=\langle \psi(t) \vert V \vert \psi(t) \rangle$, namely
\begin{equation}
E_{int}(t)=E_{tot}-\big(E_1(t)+E_2(t)\big) \,.
\end{equation}
From the left frame of Figure~\ref{fig:energy_small}, for this value of $\l=1$,  we find that $E_1$, $E_2$ and $E_1+E_2$ are much smaller than $E_{tot}\simeq 0.014$, therefore $E_{int}$ is quite constant (see also comments below for the large potential). 

In the right frame of Figure~\ref{fig:energy_small},  we show  $E_1(t)$ for several values of  $\lambda$ up to $\lambda=300$, confirming the oscillating behavior ($E_2(t)$ has a similar behavior).

%%%%%%%%%%%%%%%%%%%%%%%%%
%%%%%%%%%%%%%%%%%%%%%%%%%%%

\subsection{Coupled ghost system: Large potential case}

We approach now the more dramatic case of the potential
\begin{equation}
    V=\frac\lambda2\,x^2y^2\,,
\end{equation}
which is the case also analyzed in \cite{Gross_2021}. 
The asymptotics of this potential is not subleading with respect to the two harmonic oscillators, thus we expect substantial changes.

\subsubsection{Perturbative approach}

Let us start by the perturbative analysis. Choosing again the decoupled basis, whose index is labeled by the two integers $i\equiv (n_1,n_2)$, we have the following explicit  matrix elements of the full Hamiltonian: 
\begin{align}
 \label{eq:Hn}
 \langle n_1,n_2|& H_{tot}|m_1,m_2\rangle = (\omega_1 n_1 -\omega_2 n_2)\delta_{n_1,m_1}\delta_{n_2,m_2}
    +{}\\[1ex]
    &{}+ \tilde\lambda
    \left[ \delta_{n_1,m_1+2}\sqrt{n_1(n_1+1)}+\delta_{n_1,m_1-2}\sqrt{m_1(m_1+1)}+\delta_{n_1,m_1}(2n_1+1)\right]%\nonumber \\
    \times \Big[ 1\to 2\Big]\,,\nonumber
\end{align}
where for this potential we defined $\tilde\lambda=\lambda/8\omega_1\omega_2$. 

\pagebreak[3]

The matrix corresponding to the interaction potential is the tensor product of two ``banded" matrices, each with nonvanishing entries for $n-m=-2,0,2$. 
In perturbation theory, the banded structure implies that at any given perturbative order only a finite number of terms enter the sum on intermediate states. In this sense, the problem is thus much more tractable with respect to the previous non-polynomial small potential.   

We are able to compute $E_{(0,0)}$ to the first twelve orders:
\begin{equation}
    \label{eq:e0pert}
    \begin{split}
    E_{(0,0)}&=
    \frac{\lambda }{8 \sqrt{2}}+\frac{\left(2+5 \sqrt{2}\right)}{256}  \lambda ^2+\frac{3 \left(28+25
   \sqrt{2}\right) \lambda ^3}{2048}+\frac{\left(96326+73729 \sqrt{2}\right) \lambda
   ^4}{917504}+\ldots
   %+{}\\ 
   %&\phantom{=}+\frac{\left(75329028+54587249 \sqrt{2}\right) \lambda^5}{205520896}+\frac{\left(893685638918+638270215265 \sqrt{2}\right) \lambda^6}{782623571968}
   %   +O\left(\lambda ^7\right)
   \\[1ex]
   &\simeq 
   0.088 \,\lambda 
   +0.035 \,\lambda ^2
   +0.093\, \lambda ^3 
   +0.22\, \lambda ^4
   +0.76\, \lambda ^5
   +2.4\, \lambda ^6
   +11.\, \lambda ^7 +{}\\[.7ex]
   &\hphantom{\simeq} 
   + 43.\, \lambda ^8
   +379.\,\lambda^9
   +2.3\,\times 10^4 \lambda^{10}
   +1.2\,\times 10^5\lambda^{11}
   +3.0\,\times 10^6\lambda^{12}
   +O\left(\lambda ^{13}\right),
   \end{split}
\end{equation}
where the first order correction is indeed $\tilde\l=\l/8\sqrt{2}\simeq 0.088 \l$.

This expression explicitly confirms the expected asymptotic nature of the expansion,  with the truncation being optimal up to $\lambda\lesssim 0.3$ at fourth order, but reducing to $\lambda\lesssim 0.1$ at higher orders. Despite the asymptotic nature of the perturbative series, one expects the first perturbative orders to be a good approximation to the \emph{low-lying} energy levels,  such as $\bar E_{(0,0)}$, which are supported in a small enough region where the perturbing potential is subleading.

Some insight from the perturbative diagonalization regards also the eigenvectors and suggests that the decay is suppressed for small $\lambda$. In fact, due to the banded structure of the biquadratic potential, an estimate for the perturbative diagonalizing matrix is,
\begin{equation}
 \langle v_{(n_1,n_2)} \,|\, \bar v_{(m_1,m_2)} \rangle \propto \lambda^{\max\left[\frac{|n_1-m_1|}{2},\frac{|n_2-m_2|}{2}\right]} 
 \qquad \text{(and 0 if any $n_i-m_i$ is odd)}\,.
 \end{equation}
Therefore, the mixing between  states with distant $(n_1,n_2)$ is suppressed by high powers of $\lambda$. 

There will appear  small denominators in the perturbative expansion due to nearby (unperturbed)  eigenvalues, but this enhancement is of order $O(n)$ and thus not enough to compensate the powers of $\lambda^n$. 
For  small $\lambda$, this implies that the superposition of the would-be ground state with another nearby energy state state having very large $n_{1,2}$ is highly suppressed, thus limiting the decay.  In other words, even though the spectrum is effectively continuous, the decay in time is suppressed.

This hint is indeed confirmed quantitatively via the numerical analyses to which we now turn. 
Due to the poorer convergence, we will approach this large potential case with two numerical methods. A discretization as in the previous case, and a diagonalization in the unperturbed basis. At the end, we will cross check the compatibility of results of the two schemes.

\subsubsection{First numerical approach: Discretization}\label{subsec:discr:large}

We follow the same approach described in section~\ref{subsec:discr:small}, with the only difference that the potential part of discretized Hamiltonian now reads
\begin{equation}
\label{eq:discrete_large_pot}
H_{pot}\, \psi_{ak,aj}  = \left[\frac{1}{2} \omega _1^2 \left(a^2 j^2-\frac{1}{\omega _1}\right) -\frac{1}{2} \omega _2^2 \left(a^2
   k^2-\frac{1}{\omega _2}\right)  + \frac{\lambda}{2}a^4 j^2 k^2  \right] \psi _{a k,a j} \,,
\end{equation}
where, consistently with \eqref{mainH}, we have again subtracted $1/\omega_{1,2}$. The last term couples the two oscillators.

%%%%%%%%%%%%%%%%%%%%%%%%%
 \begin{figure}[t]
 \centerline{%
\includegraphics[width=.99\columnwidth]{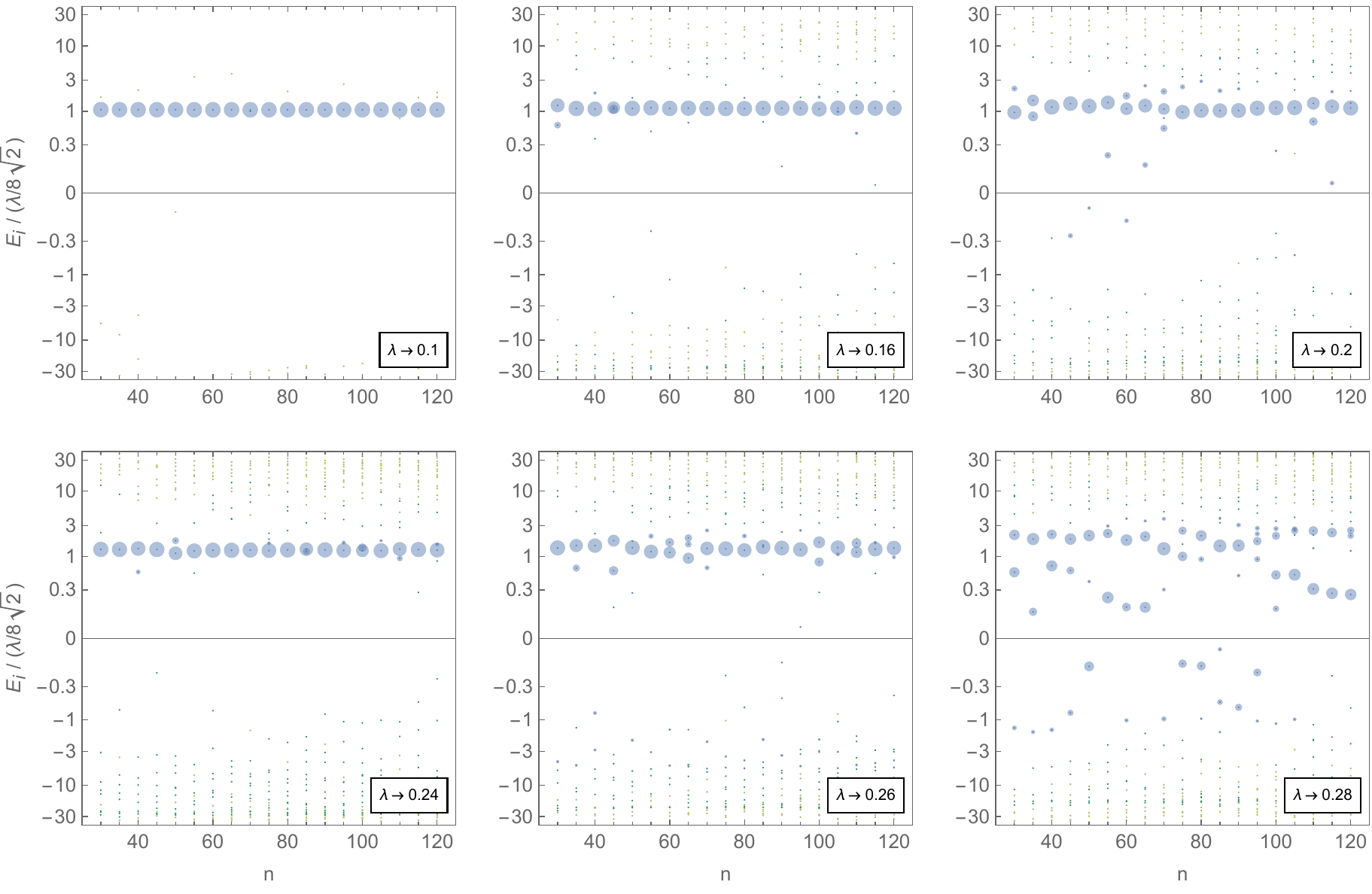}%
}
\vspace*{-2ex}
\caption{Eigenvalues for various $n$ and $\l$, the dot size showing the overlap $c_i^{(n)}$ with the initial state. The majority of eigenvalues, having negligible overlap, are still shown with minimal dot size. The eigenvalues are divided by the first order perturbative value, to test the agreement with perturbation theory. For the larger values of $\l$, we note that one can not achieve convergence of the $c_i^{(n)}$.}
\label{fig:pallini3}
\end{figure}
%%%%%%%%%%%%%%%%%%%%%%%%%%%

%%%%%%%%%%%%%%%%%%%%%%%%%
 \begin{figure}[t]
 \centerline{
\includegraphics[width=.5\columnwidth]{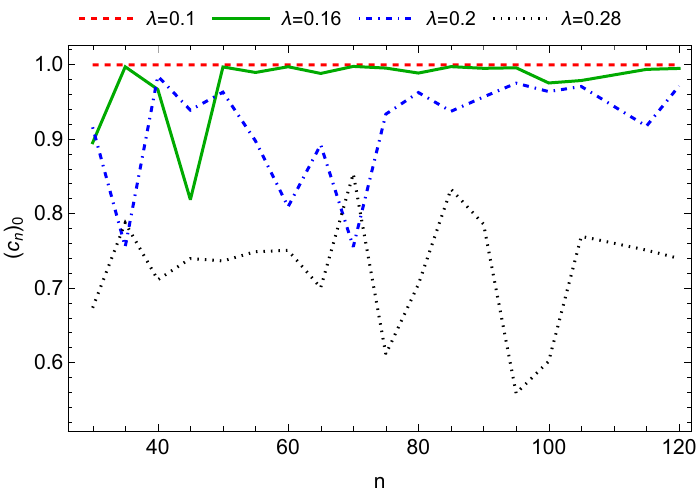}
}
\vspace*{-2ex}
\caption{(Non)convergence of the largest overlap coefficient   with the initial state $c_0^{(n)}$, for increasing $n$.  Convergence is apparent for small $\lambda=0.1$ (red dashed) or partially $\lambda=0.16$ (green solid), while it  is progressively worse and absent as $\lambda$ grows.}
\label{plotweakH}
\end{figure}
%%%%%%%%%%%%%%%%%%%%%%%%%%%

The full Hamiltonian summing the kinetic  \eqref{eq:discrete_kin} and potential terms \eqref{eq:discrete_large_pot} is diagonalized numerically, for increasing values of $n$, to yield eigenvalues $E_i^{(n)}$ and states $|v_i^{(n)}\rangle$ of the full system. As above, we compute also the would-be ground state $|\bar v_0^{(n)}\rangle$ in the free system, and thus  the overlap coefficients $c_i^{(n)}$. 

\medskip

%%%%%%%%%%%%%%%%%%%%%%%%%
 \begin{figure}[t]
 \centerline{
\includegraphics[width=.4\columnwidth]{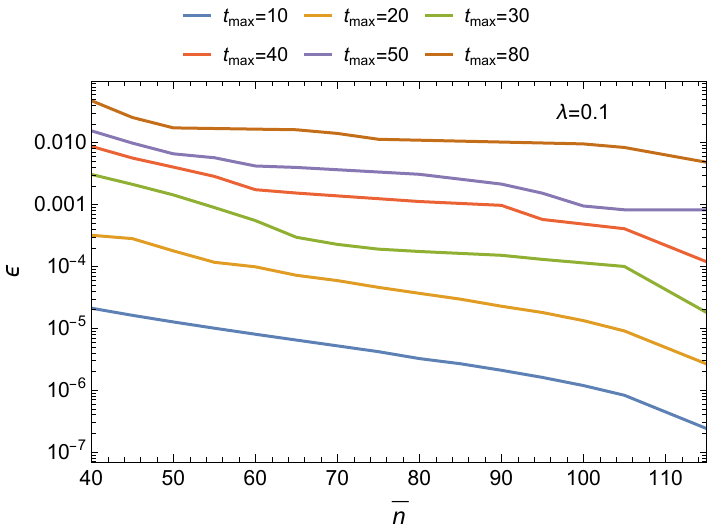}\quad
\includegraphics[width=.4\columnwidth]{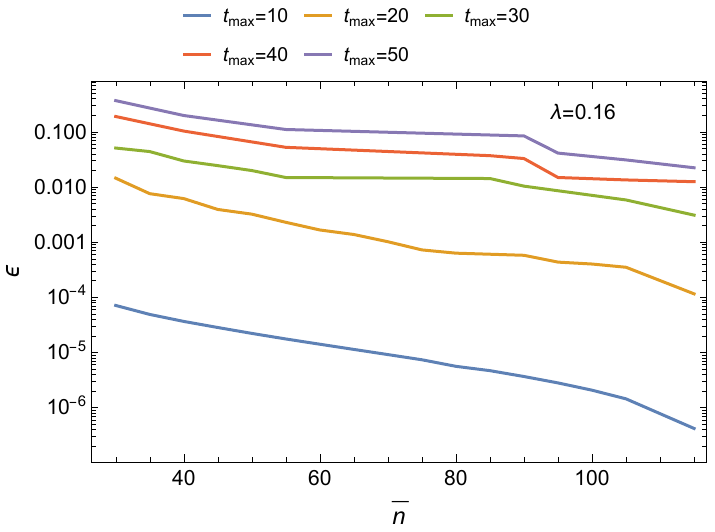}
}
\centerline{
\includegraphics[width=.4\columnwidth]{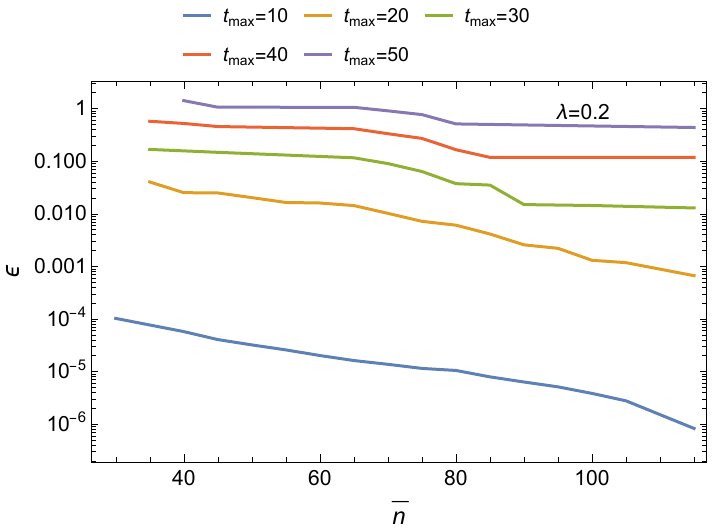}\quad
\includegraphics[width=.4\columnwidth]{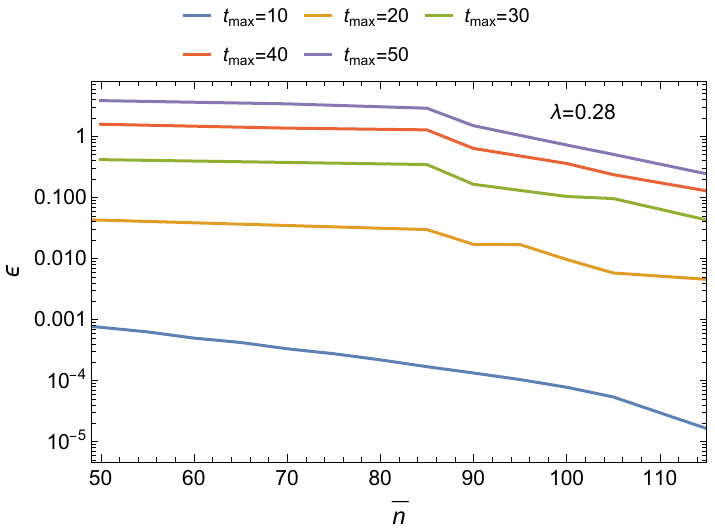}
}
\vspace*{-2ex}
\caption{Illustration of \eqref{meanconv} for $\lambda=0.1, 0.16, 0.2, 0.28$ with three different values of $t_{max}$.}
\label{plotmeanconv}
\end{figure}
%%%%%%%%%%%%%%%%%%%%%%%%%%%

Let us discuss our findings. 
\begin{itemize}

\item The numerical eigenvalues $E_i$ with overlap coefficients $c_i^{(n)}$ defined in \eqref{CC} are displayed in Figure~\ref{fig:pallini3}, for various values of $\l$ and a sequence of $n$ up to the maximal possible $n=120$. The dot size shows the magnitude of the overlap coefficients $c_i$.  Thus one can see that for low $\l$ the maximal overlap is with a state with energy equal to the perturbative first order, i.e., with the perturbative modification of the would-be ground state. This pattern starts to break beyond $\l\simeq 0.2$. Especially beyond 0.24-0.28 many states can share a substantial overlap. One notices also that the pattern is not stable by varying $n$. 

\item  
Specifically, the largest overlap coefficient
$c^{(n)}_0$ as a function of $n$ is shown additionally in Figure~\ref{plotweakH}. We can claim  convergence of $c_0^{(n)}$ only for very small $\lambda$, while already for $\l\gtrsim 0.2$ the coefficients do not display a convergent behavior. Thus one can not test the convergence of the $c_i$ as in \eqref{weakH}, for arbitrary $\lambda$.

\item On the contrary, we find numerical convergence of $P_n(t)$,  for increasing cutoff $n$, which holds  up to some time,  growing with $n$. For example, using a criterion of integrated convergence on a time interval, 
\begin{equation}\label{meanconv}
\forall\, t_{max}, \epsilon,\quad  \exists \, \bar n \quad \vert\quad\int_0^{t_{max}} \! \left| P_n(t)-P_m(t)  \right| dt \leq \epsilon \quad \forall \, n,m \geq \bar{n}  \,,
\end{equation}
we find the convergence pattern shown in Figure~\ref{plotmeanconv}: for fixed $t_{max}$, one can indeed reach smaller $\epsilon$ by increasing $\bar n$.  Alternatively, a larger $t_{max}$ requires  an increase of $\epsilon$ or a large increase of $n$.  Therefore, given our limited maximal $n$, we can estimate the maximum time up to which one can trust $P_n(t)$, within a desired precision $\epsilon$. In other words, \eqref{meanconv} provides a rationale for estimating the time within which the $P_n(t)$ can be trusted, for a choice of $\epsilon$.

%%%%%%%%%%%%%%%%%%%%%%%%%
 \begin{figure}[b]
 \centerline{
\includegraphics[width=.57\columnwidth]{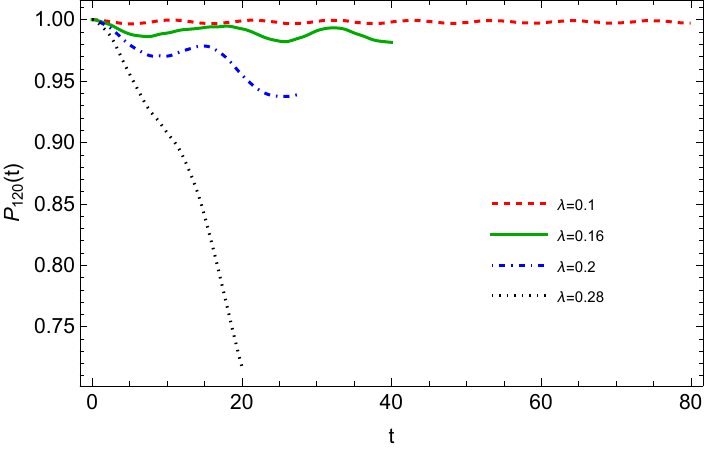}
}%
\vspace*{-2ex}
\caption{Survival probability of the would-be ground state ($n=120$).}
\label{fig:Prob}
\end{figure}
%%%%%%%%%%%%%%%%%%%%%%%%%%%

\item The actual functions $P_n(t)$ are shown in Figure~\ref{fig:Prob}. We fix the maximal time in which the probabilities are drawn, by conservatively fixing the maximal deviation  in \eqref{meanconv} as $\epsilon\leq 1\%$ at $\bar{n}=100$. This determines a validity cut for different times and for $\lambda=0.1$, 0.16, 0.2, 0.28, respectively. As we see, the validity worsens as $\l$ grows, while, for value $\lambda\ll0.1$ not shown in the plot, it improves of orders of magnitude.

%%%%%%%%%%%%%%%%%%%%%%%%%
 \begin{figure}[t]
 \centerline{%
\includegraphics[height=.313\columnwidth]{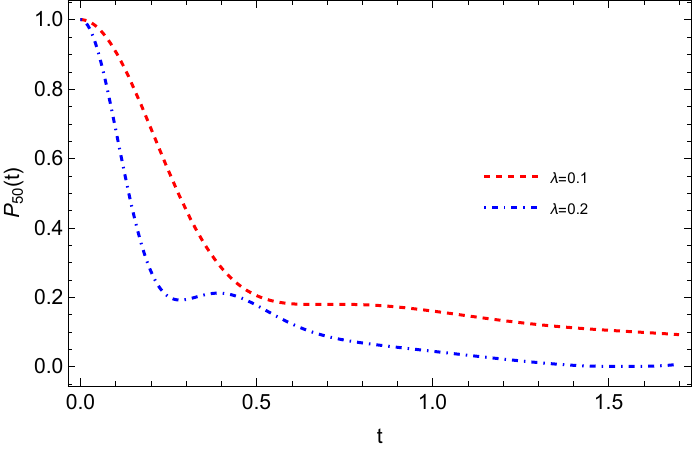}%
\quad
\includegraphics[height=.32\columnwidth]{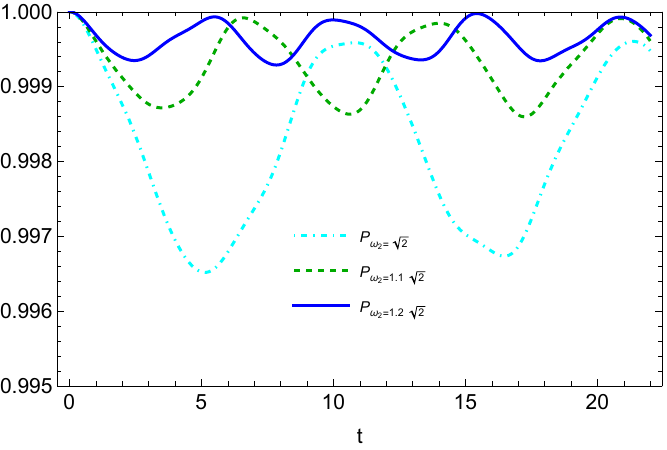}
}%
\vspace*{-2ex}
\caption{Left: The survival probabilities of excited states, here $E_1\simeq E_2 \simeq 10$, decays to zero unlike the case of the would-be ground state.  Right: Survival of the would-be ground state for increased values of  $\omega_2$ (ghost). The choice $\lambda=0.1$ ensures better convergence of $P$ and allows even a small $n=40$.  For larger $\omega_2$,  $P(t)$ oscillates closer to 1, as it is expected by the decoupling limit. }
\label{Probhigh}
\label{Figdiffomega}
\end{figure}
%%%%%%%%%%%%%%%%%%%%%%%%%%%

\item The original stability question can now be addressed. By observing the survival probability $P_n(t)$,  we note that there is no decay to zero of the would-be ground state, at least when the parameter $\lambda$ is sufficiently small. Beside a mild reduction of the survival probability, we find some oscillating behavior, well within the reliability time. This is evident for $\l=0.1$ or $\l= 0.16$.  For larger $\l$ the reliability time is not sufficient to draw a conclusion, and we will return to this important point with the other numerical scheme, below.

\item Before passing to the other regularization scheme, it is also interesting to address the destiny of the "excited" states. For illustration, we choose an initial state where the two oscillators still have balanced energies, but quite high in magnitude,  $E_1\simeq |E_2|=\simeq  10$. The result is shown in Figure\ \ref{Probhigh}. In this case, we see that the decay is much more pronounced: for $\lambda=0.1$, $0.2$, the survival probabilities decay faster and to much lower values than those of the would-be ground state. As in the previous analyses, the cut on time is performed where the functions $P_n(t)$ depart unacceptably from each other for varying $n$, signaling a cutoff dependence.

\item 
Finally, we also address the dependence on the ratio of $\w_1$ and $\w_2$. We increase $\omega_2$---the frequency of the ghost subsystem---to test the intuition of decoupling: the expectation is that in the limit of $\omega_2 \rightarrow \infty$ the ghost should decouple, and the system should be stable, i.e.,\ $P(t)\to 1$.   Quantitatively, this is confirmed in Figure~\ref{Figdiffomega}, where we raise $\omega_2$, with fixed $\omega_1=1$. We  choose here $\lambda=0.1$ since this value guarantees longer term numerical reliability. As one can see, the larger is the ratio $\omega_2/\omega_1$ the closer to one is the survival probability of the would-be ground state.

\end{itemize}

Summarizing this section, Figure~\ref{fig:Prob} shows definitely that the survival probability of the would-be ground state does not exhibit a decaying behavior, at least for a small coupling, $\lambda \lesssim 0.16$.  What the analysis uncovers is a residual oscillating behavior. 
For progressively larger values of $\l$, the $P(t)$ start to agree with a more decaying behavior. Clearly, one cannot draw a definite conclusion in those cases because of the short reliability time window.  To improve these cuts, we would need to increase $n$ (specifically, $n>120$), with a remarkable requirement of computational resources: from \eqref{Dim}, one sees that the matrix to diagonalize, for $n=120$, has $\approx 3.\times 10^9$ entries.  We will then address the same problem and these issues in the second numerical scheme.

\subsubsection{Second numerical approach: Unperturbed basis.}

The above discretization of the Schroedinger equation implies quite heavy and non-sparse numerical matrices that limit the maximal $n$ which can be reasonably diagonalized, to circa 100.  A second approach, that avoids discretization, consists in  adopting the known  basis $|\bar v_i\rangle$ of the decoupled case, and diagonalizing the full  Hamiltonian~\eqref{eq:Hn}.  

Again, the numerical scheme requires a truncation of the Hilbert space to finite dimension, thus introducing  a UV cutoff. 
This approach has formally no IR cutoff, but since the harmonic oscillator states are suppressed also in space  beyond a distance  $\sim \sqrt{n}$, this approach partially parallels the discretization cutoff~\eqref{cutoff} where both momentum and volume cutoffs grow as $\sqrt{n}$.

\begin{figure}[b]
 \centerline{
\includegraphics[width=.9\columnwidth]{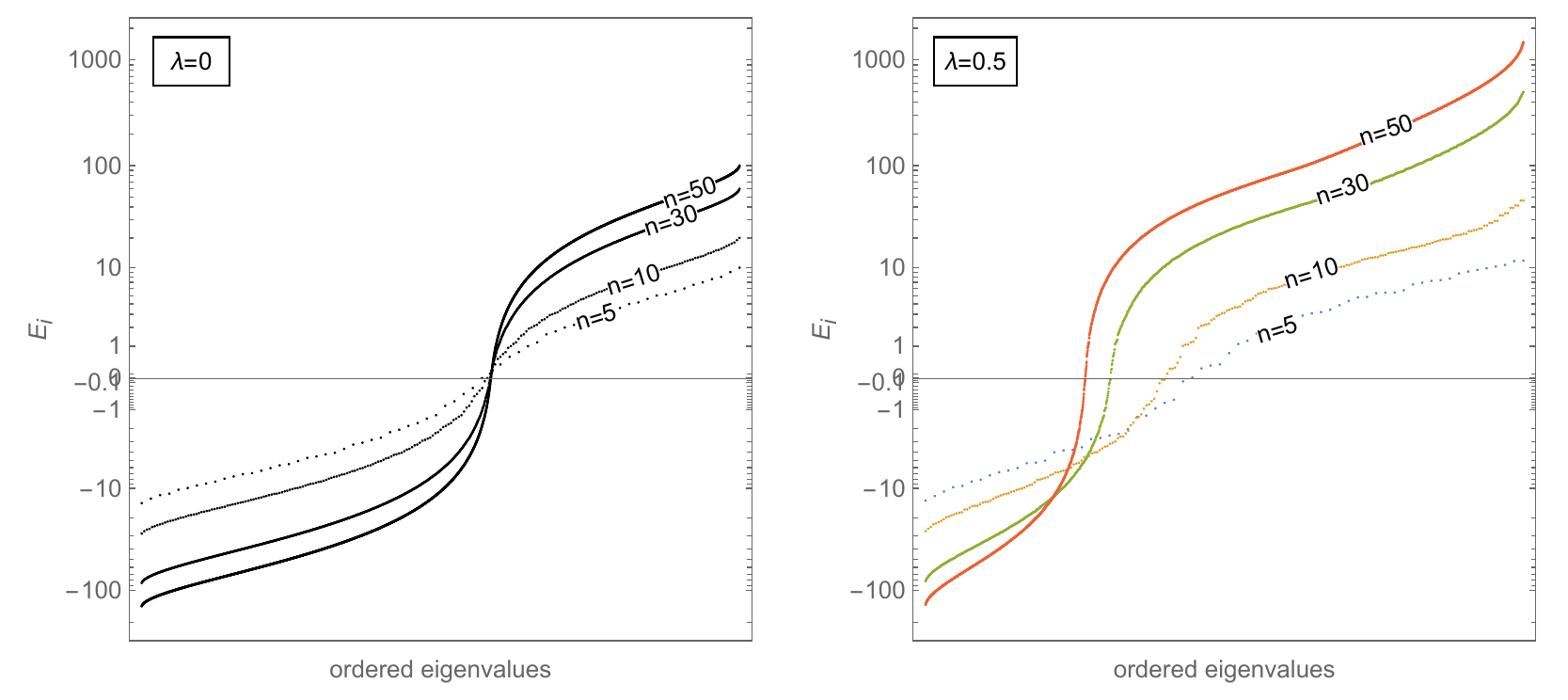}%~~
}
\vspace*{-2ex}
\caption{Spectrum: ordered eigenvalues for the decoupled case and for  $\lambda=0.5$, for various values of $n$. The "$\sinh$" scale on the vertical axis zooms around the zero energy, showing the increase in density.}
\label{fig:spectrum}
\end{figure}

It is convenient to introduce the cutoff $n$ as limiting the maximal $n_1+n_2$. Then, we can order the triangular truncated basis into a 1D list, with increasing $s=n_1+n_2 \in [0,2n]$ where for each $s$ we span all $d=n_1-n_2\in[-s,s]$ in steps of two.  For instance, we have for $n=2$,
\begin{equation}
\label{eq:biindex}
n=2\qquad \to\qquad i\in\left(\begin{array}{ccccccccccccccc}
 0 & 0 & 1 & 0 & 1 & 2 & 0 & 1 & 2 & 3 & 0 & 1 & 2 & 3 & 4 \\
 0 & 1 & 0 & 2 & 1 & 0 & 3 & 2 & 1 & 0 & 4 & 3 & 2 & 1 & 0 \\
\end{array}\right)
\end{equation}
For a given $n$, the number of elements in the basis are  $(2n+1)(n+1)\sim 2n^2$.
Thus, the truncated Hamiltonian matrix has $\sim 4n^4 $ entries, and it is also a sparse matrix  of banded form, with the number of diagonals increasing with the maximal $n$. Again, we have   for, e.g.,\ $n=2$ 
\def\tl{\tilde\lambda}
\begin{equation}
H_{tot}^{(2)}=\scalebox{0.47}{$
\renewcommand\arraystretch{1.5}\left(
\begin{array}{ccccccccccccccc}
 \tilde{\lambda } & 0 & 0 & \sqrt{2} \tilde{\lambda } & 0 & \sqrt{2} \tilde{\lambda } & 0 & 0 & 0 & 0 & 0 & 0 &
   2 \tilde{\lambda } & 0 & 0 \\
 0 & 3 \tilde{\lambda }-\omega _2\!\! & 0 & 0 & 0 & 0 & \sqrt{6} \tilde{\lambda } & 0 & 3 \sqrt{2} \tilde{\lambda }
   & 0 & 0 & 0 & 0 & 0 & 0 \\
 0 & 0 & \!\!3 \tilde{\lambda }+\omega _1\!\! & 0 & 0 & 0 & 0 & 3 \sqrt{2} \tilde{\lambda } & 0 & \sqrt{6}
   \tilde{\lambda } & 0 & 0 & 0 & 0 & 0 \\
 \sqrt{2} \tilde{\lambda } & 0 & 0 & \!\!5 \tilde{\lambda }-2 \omega _2 \!\!& 0 & 2 \tilde{\lambda } & 0 & 0 & 0 & 0 & 2
   \sqrt{3} \tilde{\lambda } & 0 & 5 \sqrt{2} \tilde{\lambda } & 0 & 0 \\
 0 & 0 & 0 & 0 & \!\!9 \tilde{\lambda }+\omega _1-\omega _2 \!\!& 0 & 0 & 0 & 0 & 0 & 0 & 3 \sqrt{6} \tilde{\lambda } &
   0 & 3 \sqrt{6} \tilde{\lambda } & 0 \\
 \sqrt{2} \tilde{\lambda } & 0 & 0 & 2 \tilde{\lambda } & 0 & \!\!5 \tilde{\lambda }+2 \omega _1\!\! & 0 & 0 & 0 & 0 & 0
   & 0 & 5 \sqrt{2} \tilde{\lambda } & 0 & 2 \sqrt{3} \tilde{\lambda } \\
 0 & \sqrt{6} \tilde{\lambda } & 0 & 0 & 0 & 0 & \!\!7 \tilde{\lambda }-3 \omega _2\!\! & 0 & 2 \sqrt{3} \tilde{\lambda
   } & 0 & 0 & 0 & 0 & 0 & 0 \\
 0 & 0 & 3 \sqrt{2} \tilde{\lambda } & 0 & 0 & 0 & 0 & \!\! 15 \tilde{\lambda }+\omega _1-2 \omega _2 \!\! & 0 & 2
   \sqrt{3} \tilde{\lambda } & 0 & 0 & 0 & 0 & 0 \\
 0 & 3 \sqrt{2} \tilde{\lambda } & 0 & 0 & 0 & 0 & 2 \sqrt{3} \tilde{\lambda } & 0 & \!\!15 \tilde{\lambda }+2
   \omega _1-\omega _2\!\! & 0 & 0 & 0 & 0 & 0 & 0 \\
 0 & 0 & \sqrt{6} \tilde{\lambda } & 0 & 0 & 0 & 0 & 2 \sqrt{3} \tilde{\lambda } & 0 & \!\!7 \tilde{\lambda }+3
   \omega _1 \!\! & 0 & 0 & 0 & 0 & 0 \\
 0 & 0 & 0 & 2 \sqrt{3} \tilde{\lambda } & 0 & 0 & 0 & 0 & 0 & 0 & \!\!9 \tilde{\lambda }-4 \omega _2\!\! & 0 & 2
   \sqrt{6} \tilde{\lambda } & 0 & 0 \\
 0 & 0 & 0 & 0 & 3 \sqrt{6} \tilde{\lambda } & 0 & 0 & 0 & 0 & 0 & 0 & \!\!21 \tilde{\lambda }+\omega _1-3 \omega _2\!\!
   & 0 & 6 \tilde{\lambda } & 0 \\
 2 \tilde{\lambda } & 0 & 0 & 5 \sqrt{2} \tilde{\lambda } & 0 & 5 \sqrt{2} \tilde{\lambda } & 0 & 0 & 0 & 0 & 2
   \sqrt{6} \tilde{\lambda } & 0 & \!\!25 \tilde{\lambda }+2 \omega _1-2 \omega _2 \!\!& 0 & 2 \sqrt{6} \tilde{\lambda }
   \\
 0 & 0 & 0 & 0 & 3 \sqrt{6} \tilde{\lambda } & 0 & 0 & 0 & 0 & 0 & 0 & 6 \tilde{\lambda } & 0 & \!\!21
   \tilde{\lambda }+3 \omega _1-\omega _2\!\! & 0 \\
 0 & 0 & 0 & 0 & 0 & 2 \sqrt{3} \tilde{\lambda } & 0 & 0 & 0 & 0 & 0 & 0 & 2 \sqrt{6} \tilde{\lambda } & 0 & \!\!9
   \tilde{\lambda }+4 \omega _1 \\
\end{array}
\right)
$}.
\end{equation}
The first entry corresponds to the would-be ground state, with zero unperturbed energy and $\tilde\lambda$ as first perturbative correction. Then one notices explicitly that, although some other diagonal entry may end up being nearly degenerate with zero, for a cancellation of $n_1-\sqrt{2}n_2\sim 0$,  the would be ground state can mix with that state only through a large number of off-diagonal terms, implying higher powers of $\lambda$. As a result, the decay of the would-be ground state is suppressed even if the spectrum becomes dense. 

The Hamiltonian~\eqref{eq:Hn} can be diagonalized by standard methods dealing with sparse and banded matrices. Because the entries grow as $4n^4$, the problem becomes quickly challenging: we are able to tackle numerical diagonalization up to $n\simeq 200$, i.e.\ circa 6 billion entries.

\paragraph{The spectrum.} The exact numerical eigenvalues are ordered and presented in Figure~\ref{fig:spectrum}, for the decoupled and for a coupled case. A number of results can be drawn:

\begin{figure}[t]
 \centerline{
\includegraphics[width=.48\columnwidth]{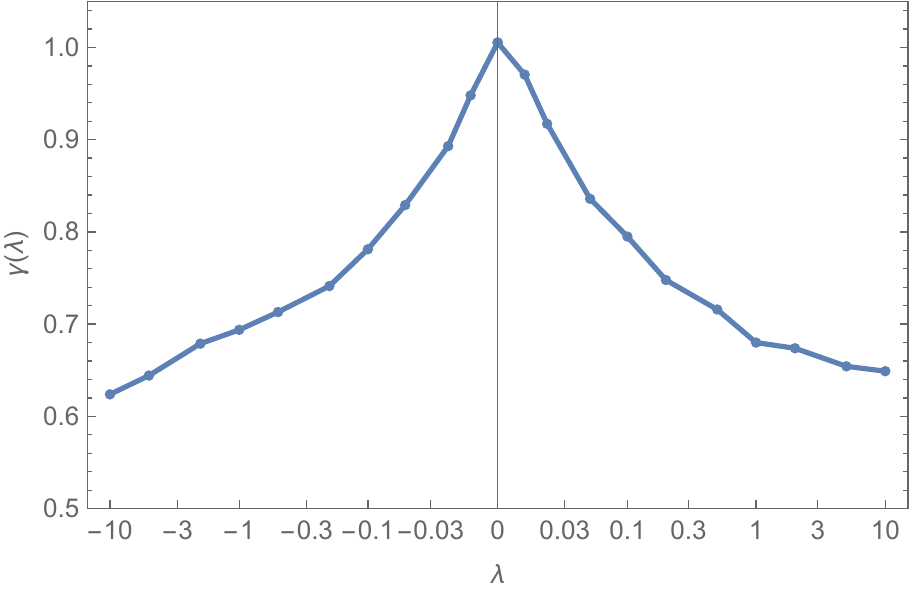}~~
\includegraphics[width=.48\columnwidth]{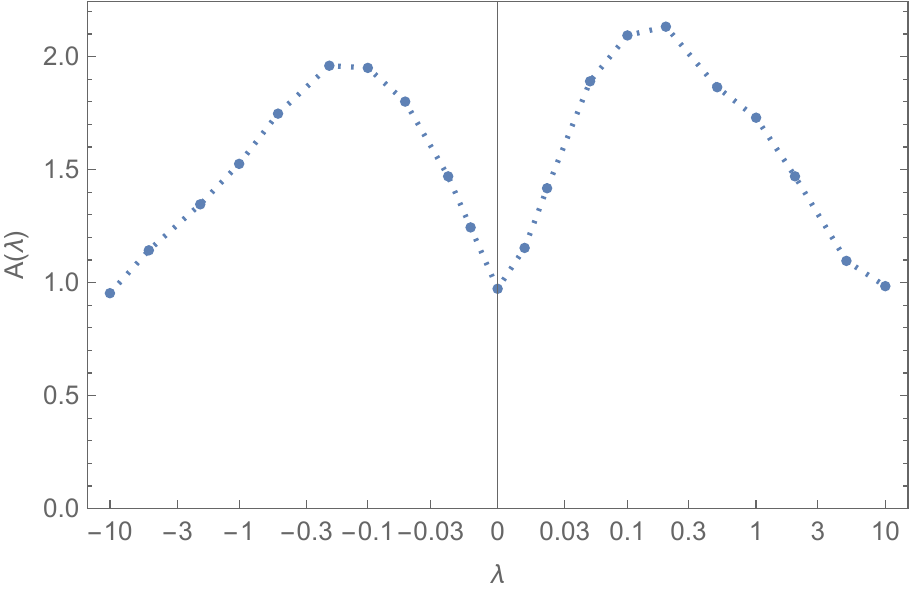}
}
\vspace*{-2ex}
\caption{Left plot: exponent  $\gamma(\l)$ for the spectral density.  Right plot: prefactor $A(\l)$. The curves are calculated for $n$ up to $250$.}
\label{fig:gamma}
\end{figure}

 \begin{figure}[t]
 \centerline{
\includegraphics[width=.999\columnwidth]{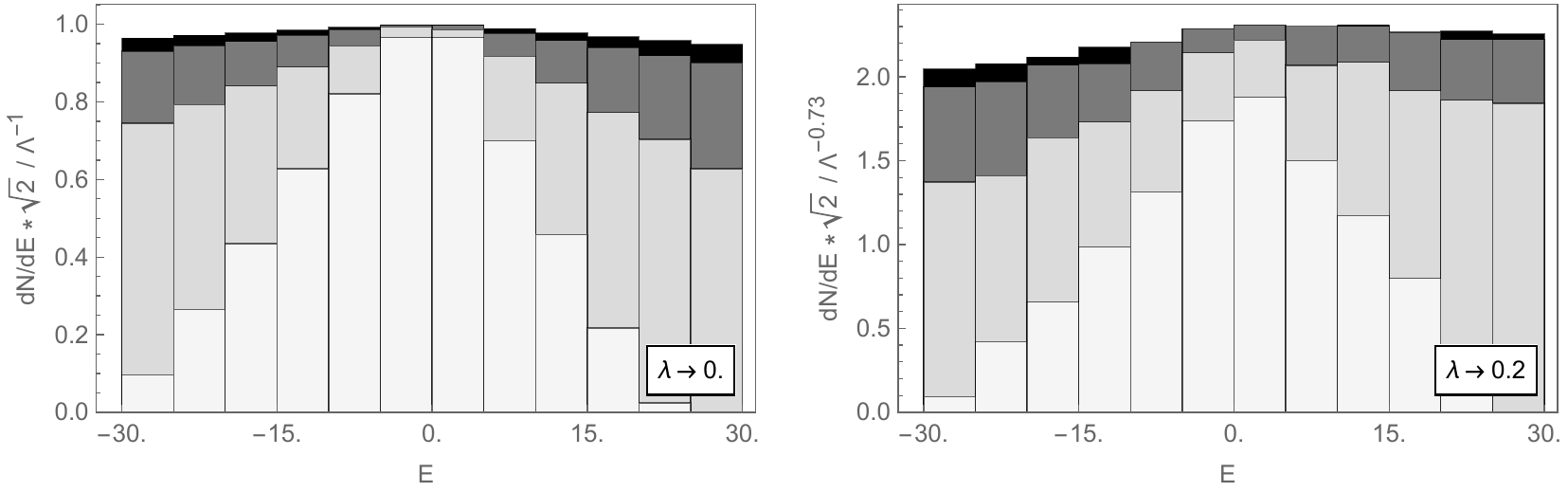}
}
\vspace*{-2ex}
\caption{Density of states histogram for $\l=0$ (left) and $\l=0.2$ (right), rescaled respectively with $\Lambda^{-1}$ and $\Lambda^{-0.73}$. The increasingly darker colors represent the spectra calculated for $n=10,35,130,250$. Like in the known  decoupled case, also in the interacting case the spectrum tends to be flat in energy for large $n$, with some deformation towards the right for positive $\lambda$.}
\label{fig:dnde}
\end{figure}

\begin{enumerate}
    \item As expected, by increasing $n$ the largest and lowest eigenvalues become larger, signaling the removal of the UV cutoff. 
    At the same time the spectrum in the large $n$ limit becomes increasingly dense, signaling the removal also of the IR cutoff. 

    In Figure~\ref{fig:spectrum} we also zoom on the  states near zero energy, and one can verify the increase in the density of states there.
    This was discussed in the decoupled case, but  we see that the same happens also for $\lambda\neq0$. 
    
    \item   
    The distribution of states in fact confirms the one in the decoupled case, with some interesting modifications for $\l\neq0$.   In the decoupled system we found that the density of states grows linearly with the UV cutoff $\Lambda\propto n$, see~\eqref{eq:rhoofE}. To compare with that result, we consider that the triangular cutoff arrangement~\eqref{eq:biindex} relates  this $n$ and the previous cutoff as  $\Lambda = 2(2- \sqrt{2})n\,\w_1\simeq 1.17\,n$. We then find that, in the UV limit, the density of states is well approximated as
\begin{equation}
\frac{{\rm d} N}{{\rm d} E} \simeq \frac{1}{\sqrt{2}}  A(\l)\,\Lambda^{\gamma(\l)}\,,
\label{eq:rhoofEfull}
\end{equation}
where the functions $A$ and  $\gamma$  encode the  dependence on $\lambda$ and tend to 1 for $\lambda\to0$, as required by the Kronecker approximation theorem. They are depicted in Figure~\ref{fig:gamma}.

What is notable is that the trend towards a dense spectrum, linear in $\Lambda$ for $\l=0$, is replaced by a smaller power of the UV cutoff, suggestive of an "anomalous" exponent. 
It is also remarkable that the  dense spectrum persists even in the strong coupling regime, here shown up to $\lambda\sim 10$.

\smallskip

At higher $\l$, since $\gamma$ is decreasing, it is tempting to speculate that the exponent $\gamma$ could approach zero, signaling a sort of UV fixed point for the spectral density.  The whole system would then retain an unbounded but \emph{discrete} spectrum.

\item We  plot the density of states, divided by $\Lambda^\gamma$ in Figure~\ref{fig:dnde}, demonstrating the trend towards a constant spectrum independent of $E$, after this rescaling.  The constant spectrum is approached with asymmetric paces, for positive and negative energies, due to the triangular cutoff used here.

\item The above results hold  for both positive and negative $\lambda$.  There is indeed no qualitative difference between these two cases, that one can relate by swapping the "normal" and "ghost" variables. Indeed  changing $\lambda\to -\lambda$ amounts to changing the overall sign of the Hamiltonian and swapping $\omega_1\leftrightarrow \omega_2$.
A mild quantitative asymmetry between the results with  positive and negative values of $\lambda$ reflects therefore just the asymmetry between $\omega_1$ and $\omega_2$.

A related comment is in order here, about the  argument by Dyson~\cite{Dyson1952} on the convergence of perturbation theory. The argument states that if a perturbative series were to be convergent for some small coupling, say $|\lambda|< R$  (some positive real number) then it should converge also for small but negative coupling. But for negative coupling the theory is usually unstable. Dyson concluded that the series can not be analytic at $\lambda=0$, rendering the expansions divergent, at best asymptotic. In our case, we want to stress that Dyson’s argument is ineffective: if the theory does not change its behavior under $\lambda\to -\lambda$, this does not automatically imply that the series is convergent. This is indeed the situation in our system: the divergence of the expansion is fundamentally tied to the lack of proper conditions for an analytic perturbation theory, as discussed in point (A) at the beginning of the section: in particular the lack of an unperturbed spectrum with isolated eigenvalues, which in turn is due to the ghost nature of one variable.
\end{enumerate}

 \begin{figure}[b]
 \centerline{
\includegraphics[width=.99\columnwidth]{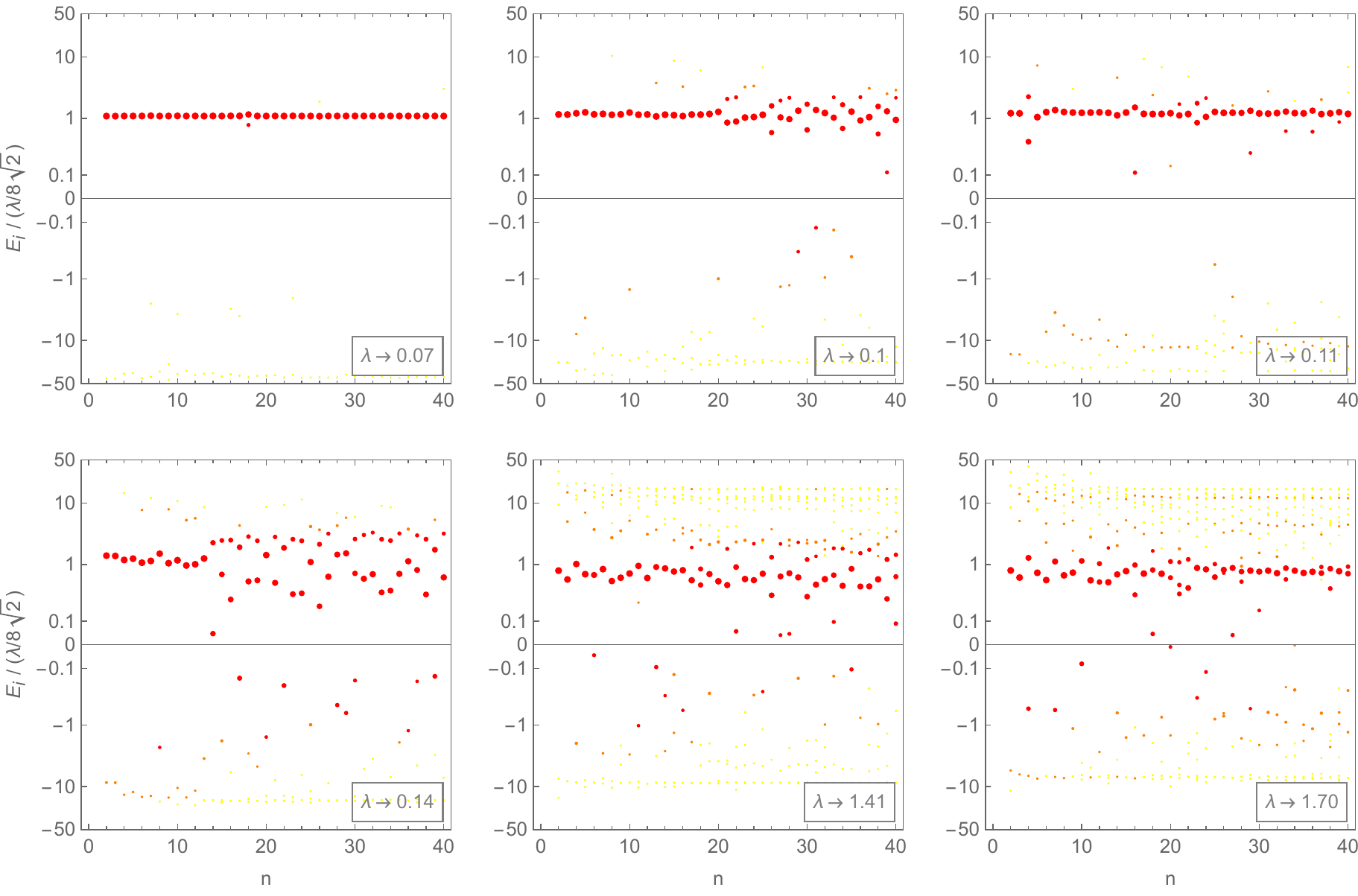}
}
\vspace*{-2ex}
\caption{Eigenvalues normalized to the first perturbative shift of the would-be ground state ($\lambda/8\sqrt{2}$), with the dot size showing the overlap with the initial state, $c_i$. The dots show points with $c_i$ larger than $10^{-1}$, $10^{-2}$, $10^{-3}$ in red, orange, yellow. Smaller overlaps are not shown.} 
\label{fig:overlap}
\end{figure}

\paragraph{Overlap integrals.}  We then study in detail the overlap integrals $c_i^{(n)}$  with the exact energy eigenstates of our initial  state $|\bar v_0\rangle=|(0,0)\rangle$. 
In Figure~\ref{fig:overlap}, we display, for each choice of $n$ up to 40, all eigenvalues placed vertically, with a dot size given by the overlaps $c_i^{(n)}$. Only the states with non-negligible overlap are shown.
The energy is also normalized to the value of the first perturbative correction to the ground state, $\tilde \lambda\equiv \lambda/8\sqrt{2}$, so that when the numerical eigenvalue matches the perturbative ground state, it  aligns with 1 on the vertical axis. 
A few comments can be drawn from those figures:
\begin{enumerate}

\item As can be seen, for small $\lambda$ (first frame) the overlap happens practically with a single eigenstate. That single state matches the perturbative value of the energy, meaning that perturbation theory is reliable.
This fact persists with large $n$; thus, even though the spectrum becomes dense due to cancellations, the initial state does not 'spill' to nearby eigenstates.  This is understandable in perturbation theory, as  states with high $(n_1, n_2)$ have very small overlap with $(0,0)$, and in the diagonalization need higher orders of $\lambda$.  

 \begin{figure}[t]
 \centerline{
\includegraphics[width=.9\columnwidth]{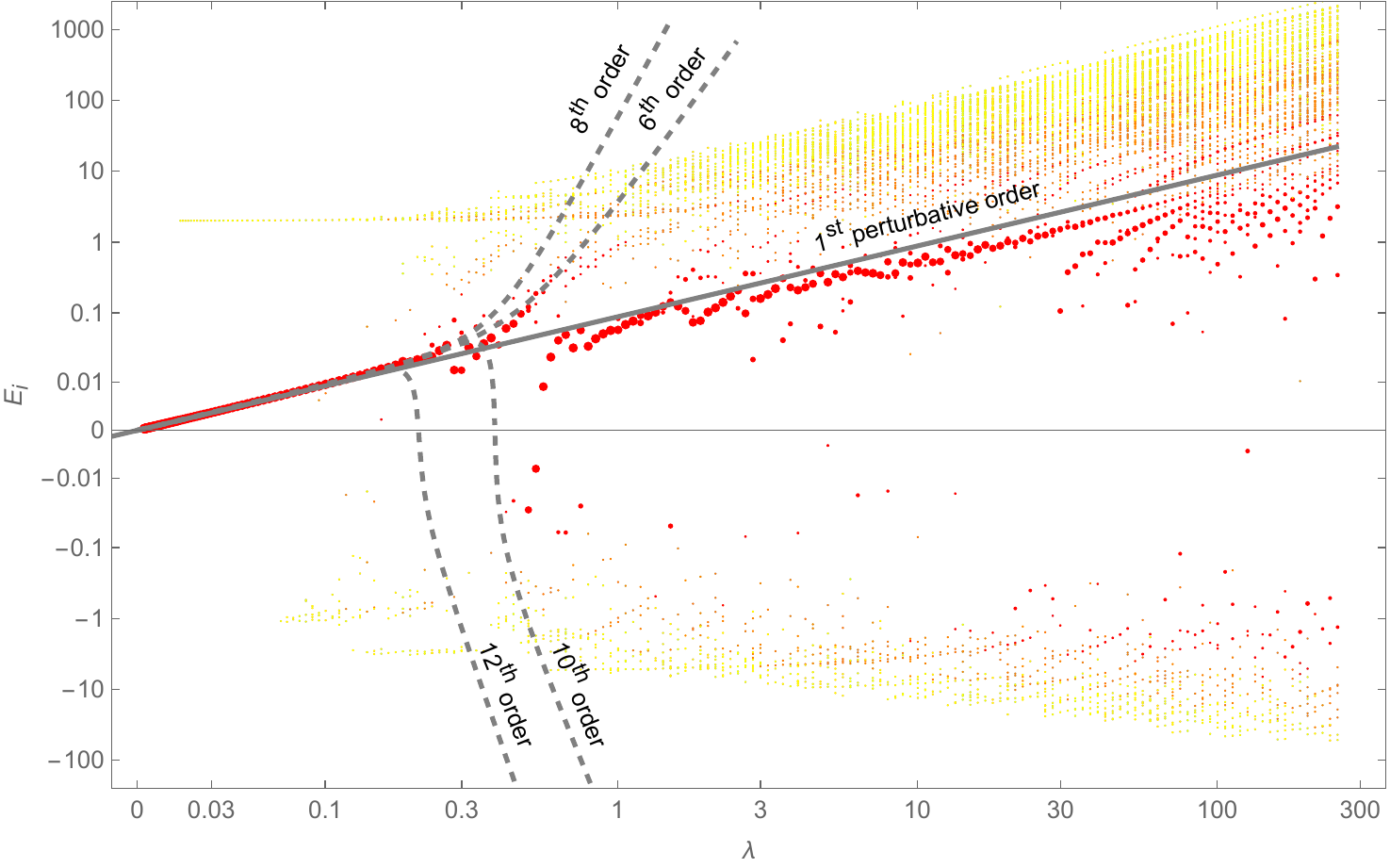}
}
\vspace*{-2ex}
\caption{Same as Figure~\ref{fig:overlap} but varying $\lambda$, for fixed $n=40$. The vertical axis represents the eigenvalues energy, and larger dot sizes represent larger overlap with the initial (0,0) state. The gray lines draws the perturbative estimate of $E_{(0,0)}$ at first and sixth order, as from \eqref{eq:e0pert}, showing that the first order follows better the states with largest overlap with the would-be ground state (0,0).} 
\label{fig:overlaplambda}
\end{figure}

\item  However, this behavior  persists qualitatively even for larger $\lambda$, and moreover the dominant overlap states lie curiously near the energy of the ground state given  by the first perturbative order.  One notices that there are  increasingly larger deviations:  
in the first frame ($\lambda=0.01$), there is some particular value of $n$, $n=18$, where the overlap is split between two states. This can be traced  to some transient eigenvalue crossing which enhances some new overlap, it appears for that accidental value of $n$, and disappears afterwards.  For larger~$\lambda$, eigenvalue crossings happen more easily, with a pattern which can be hardly analyzed. For instance, in the second and fourth frame they appear to increase with $n$, while in the third and last, the deviations subside and dominance of a single state resumes. 
Figure~\ref{fig:overlaplambda}, where we vary  $\lambda$ with $n$ fixed, shows in more detail how the patterns arise  from level crossing, where the nearby energies arise from cancellations of large positive and negative energies. It may be imagined as a manifestation of the classical chaotic nature of the system, that we will discuss in section~\ref{sec:classical}.  

\item From Figure~\ref{fig:overlaplambda},  we have confirmation that the energy of the dominantly overlapping state follows the \emph{first order} perturbed energy. This is precise in the small coupling regime $\l\lesssim 0.1$, but the same trend persists also for higher coupling, from $\l\sim 1$ and up to some highly strong coupling regime of $\lambda\sim 30$. This is a remarkable and curious fact, possibly in accord with the fact that the perturbative series is only asymptotic.

\item We finally argue that the dominance of a single overlap, likely implies that the initial state does not decay: a decaying time-evolution requires large overlap with a continuum of near eigenvalues, as understood from the Fourier analysis. 
Thus, also for large $\lambda$, where this dominance is  qualitatively holding,  we expect a survival in time of the initial state. We  address  this by studying now the survival time dependence.

\end{enumerate}

\subsubsection{Survival probability}

 With  the $E_i$ and $c_i$, we can address the time evolution of the $|0,0\rangle$ state, given by $P_n(t)$  
as in \eqref{discrete_P}. The results are in  agreement with the discretization scheme,  but we are in position to study the outcome in more detail.

 \begin{figure}[t]
 \centerline{
\includegraphics[width=.45\columnwidth]{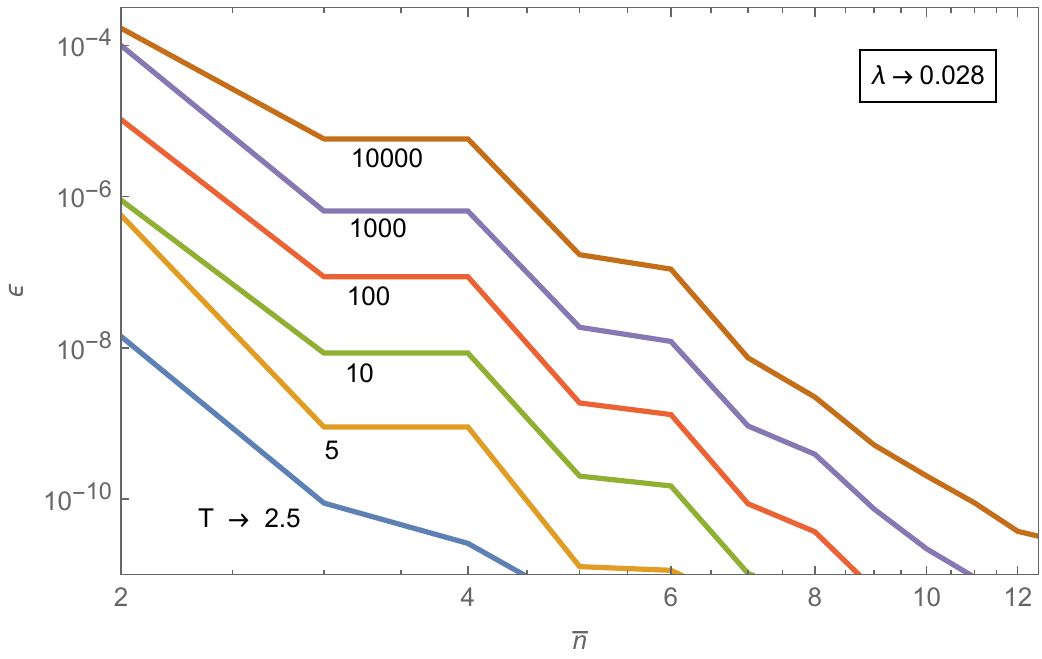}~~
\includegraphics[width=.45\columnwidth]{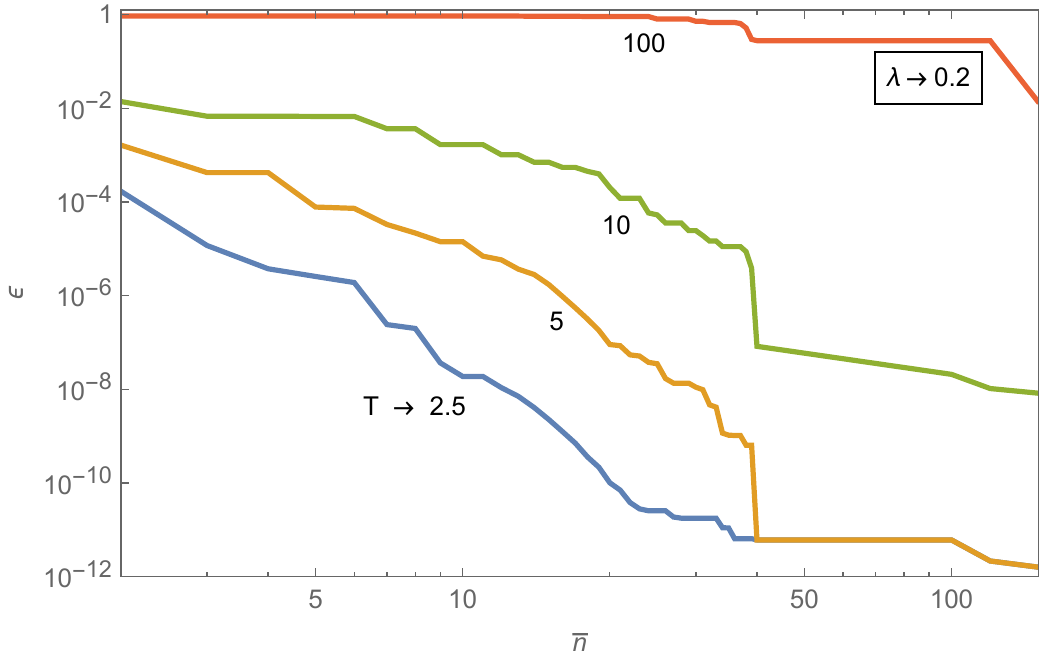}
}
\vspace*{-2ex}
\caption{Minimal $\bar n$ in order to have uniform convergence below $\epsilon$, up to $T$. Small $\l$ has very good convergence, to very long time even for small $\bar n$. Large $\l$ instead is more critical: demanding convergence up to large $T$ requires quickly growing $\bar n$, towards the computational limit (here fixed at $n=160$).} 
\label{fig:2Dconvergence}
\end{figure}

As in the discretized approach, the validity in time of the numerical solution increases with larger $n$, or by lowering $\lambda$. This is shown in Figure~\ref{fig:2Dconvergence}, where this time we  adopt   uniform (relative) convergence, namely $2|P_{n}(t)-P_{m}(t)|/ [P_{n}(t)+P_{m}(t)]<\epsilon$ for all  $n,m<\bar n$ and $0<t<T$.  
For chosen $\epsilon$ and time $T$, one finds $\bar n$ so that for higher $n$ all relative variations  are less than $\epsilon$. For instance, with $\l=0.2$, choosing $\epsilon\simeq 10^{-6}$ one has validity up to $t=2.5$ for $\bar n>5$, but to achieve validity up to $t=10$ one needs $\bar n\gtrsim 50$.  With a computational limit of $n=200$ we can test the convergence  up to $\bar n=160$. The results confirm the trend of uniform convergence:  increasing $\bar n$ allows for smaller $\epsilon$ or larger $T$. 
The trends clearly depend also on $\lambda$: we find that for small $\lambda$ one can reach much longer times (also limited by the computational precision).

 \begin{figure}[t]
 \centerline{
\includegraphics[width=.68\columnwidth]{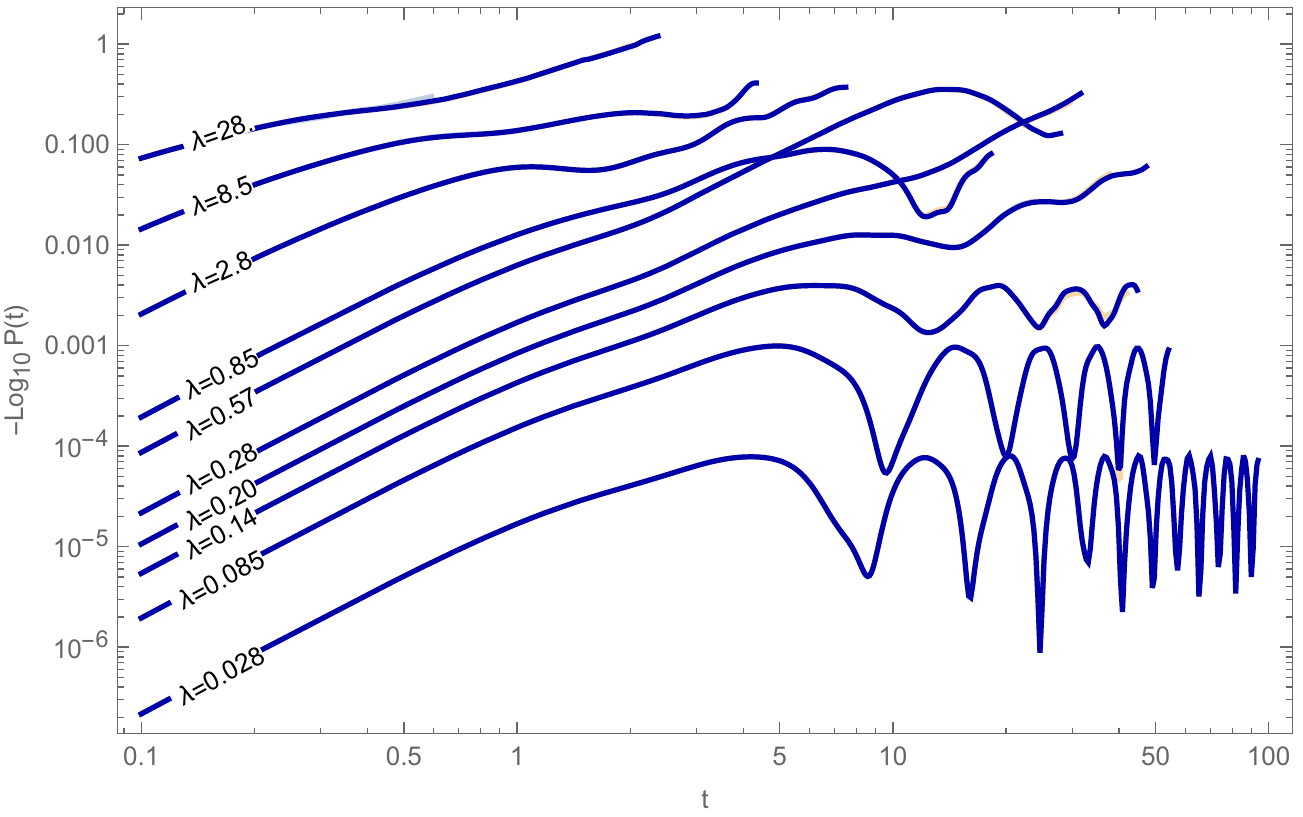}
}
\vspace*{-2ex}
\caption{Time decay of $P(t)$ for a wide range of values of $\lambda$.  To analyze the  oscillations at small $\l$, the vertical axis shows $-\log_{10}P(t)$. Each curve extends up to its reliability time, except the lowest, which would be valid up to much larger times. Curves here are calculated with a maximum $n=200$.} 
\label{fig:LOGP}
\end{figure}

For each $n$ one has thus a time range in which the $P_n(t)$ is reliable. The results are analogous to Figure~\ref{fig:Prob}, were it was found that the probability decays very little and also features an oscillatory behavior.  
It is thus useful to display the logarithm $-\log_{10} P_{(n)}(t)$, which starts from 0 and grows for decaying $P$, with 1 representing  a decay of 90\%.
The results of $P_{(n)}(t)$ for the maximum computable $n$ and reliability time $t$  are shown in Figure~\ref{fig:LOGP} for a selection of values of $\lambda$.   For the  smallest value of $\l$ we cut the plot at $t=100$ but the solution would be reliable for much longer times. 

First of all we notice that the initial decay appears to be universal, and indeed one can uncover an analytical behavior
\begin{equation}
    P(t\sim0) \sim 1-A\,  \l^2\,t^2 \,,
\end{equation}
with $A\simeq 14.$, thus a faster decay with larger $\l$. 

Then we confirm that, for small coupling, after this initial period, the decay stops and only residual oscillations appear.  The oscillations have very small amplitude (and obviously vanish for $\l\to0$) but have a typical period $\Delta t\simeq 8.$, which increases slightly with $\l$.
These trends are smooth in $\l$ although in the right plot we had to choose a discrete set of values of $\l$.  The fixed period is more visible in  previous figures with a linear time scale. 
More importantly, by analyzing  this oscillatory regime for small $\l$, one finds numerically that the survival probability remains  above a lower bound which depends on $\l$, estimated as
\begin{equation}
\label{eq:estimate}
\qquad \qquad P(t)\gtrsim 10^{-0.085\, \l^{2}-0.9 \l^3}\approx 1- 0.2\, \l^{2}-2.1\,\l^3
\,,\qquad   (\text{for }\l<0.15)
\end{equation}
thus quantifying the  (non) decay of the would-be ground state. 
As mentioned, also the time of reliability (cutoff independence) of $P$ increases non perturbatively to huge values, for vanishing $\l$. Therefore in this regime this lower bound on $P$ is robust.

Beyond $\l\simeq 0.2$  a decaying behavior seems to appear, although one can not assess whether this is just a transient decay before other oscillations, due to the limited reliability time for larger $\l$.  In addition, for particular values of $\l$, oscillations appear again, like for $\l\sim 0.85$ of Figure~\ref{fig:LOGP}, and also in many other curves not shown.

Therefore, we wish to conclude that oscillations of $P$ are a typical feature, interrupted only by some particular values of $\l$ where resonances enhance more definitely the decay, via the pattern of spectrum and overlaps coefficients seen in Figure~\ref{fig:overlaplambda}.

\begin{figure}[b]
 \centerline{
\includegraphics[height=.35\columnwidth]{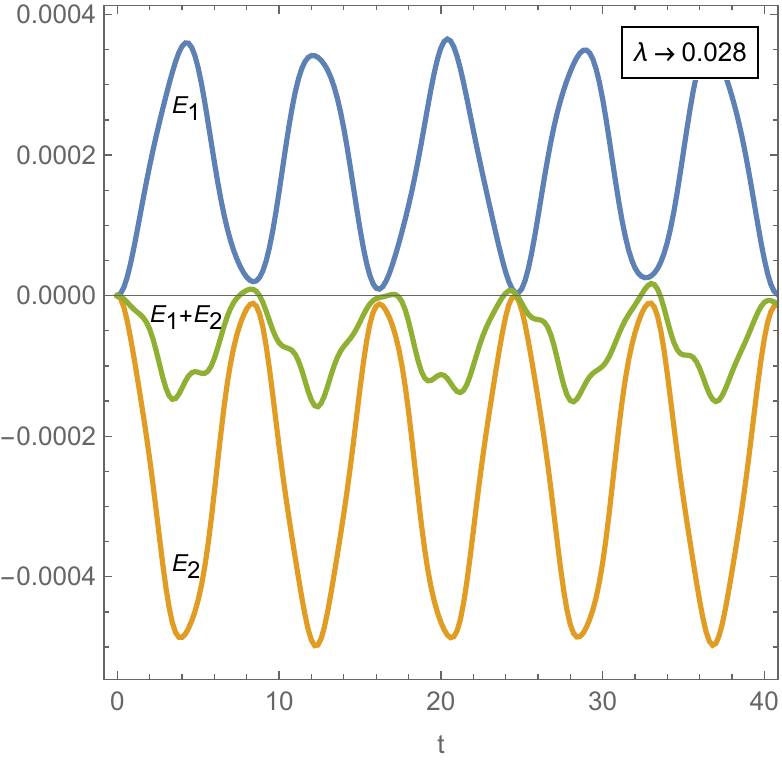}
\includegraphics[height=.35\columnwidth]{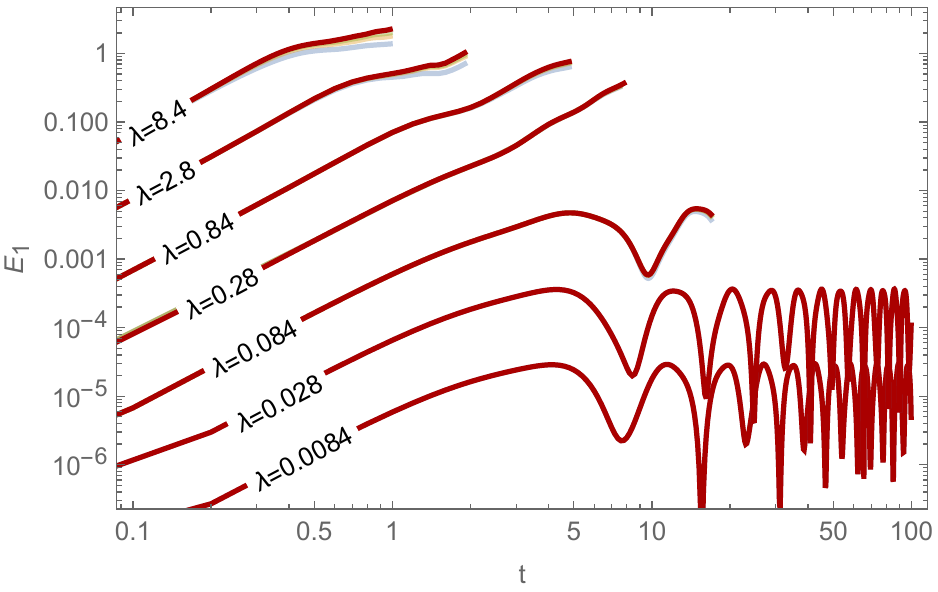}}
\vspace*{-2ex}
\caption{Left: mean energy in time for each oscillator and their sum. The total conserved energy amounts to $\tilde\l\simeq 0.002$, larger than the 
the reported energies.  Right: Energy gained by the first oscillator. Curves in these plots are calculated with a maximum of $n=100$.} 
\label{fig:energies_large}
\end{figure}

\subsubsection{Extracting energy?}
\label{sec:energy}

In Figure~\ref{fig:energies_large} we display the balance of the energies stored in the two oscillators $E_1(t)$ and $E_2(t)$ (see~\ref{eq:E1E2}) and we report also in detail the energy gained by the first oscillator for some selected values of $\l$.   The figures confirm the  picture outlined above: for small $\l$, after an initial short transient, the system settles in an oscillating behavior, where the two subsystems exchange periodically some energy, but do not show signs of runaway. This is evident for small $\l$ where the reliability time of the solutions is long.    The oscillating behavior implies that a short-time average of the energies reaches quickly a constant value after which it remains constant. 

For small $\l$, the value of the average energy gained by the first oscillator is substantially smaller than the total initial (and conserved) energy,  $E_{tot}=\langle\bar v_0| H |\bar v_0\rangle\langle\bar v_0| V_{int} |\bar v_0\rangle=\tilde\l=\l/8\sqrt{2}$. In this regime therefore, along the time evolution the total energy remains  dominated by the interacting potential energy, while the oscillator energies, positive and negative, do not grow.  This fact suggests that for  small $\l$ in this system the interaction term allows only for a limited exchange of energy, unable to destabilize the evolution.

We find numerically that $E_1$ overshoots $E_{tot}=\tilde\l$ only for approximately $\l> 0.2$, as it can be appreciated in the right frame of Figure~\ref{fig:energies_large}. This suggests that below this value the dynamics does not produce an instability but just an oscillation, while the possibility of decay opens up beyond it.

The analysis of the oscillatory regime is also of interest, because it can shed light on the way in which the system avoids a runaway fate, despite the interaction term coupling the normal and ghost variables. 
To this aim, we study briefly the probability currents in the $x$--$y$ plane.

\subsubsection{Where does the probability (not) flow?}

%%%%%%%%%%%%%%%%%%%%%%%%%
 \begin{figure}[b]
 \centerline{%
\includegraphics[width=.4\columnwidth]{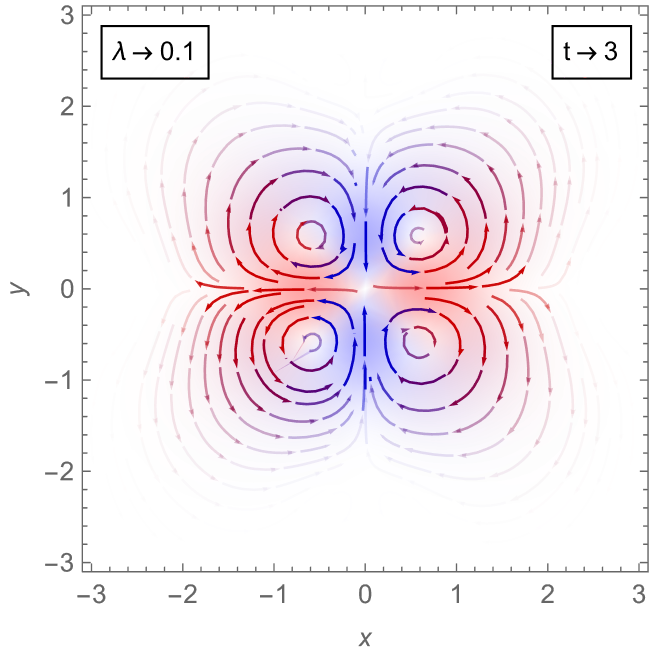}%
\includegraphics[width=.4\columnwidth]{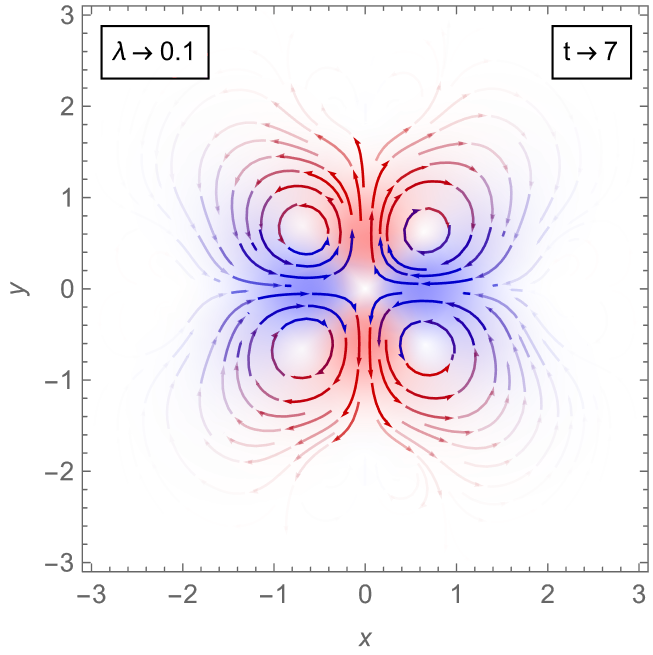}%
} \centerline{%
\includegraphics[width=.4\columnwidth]{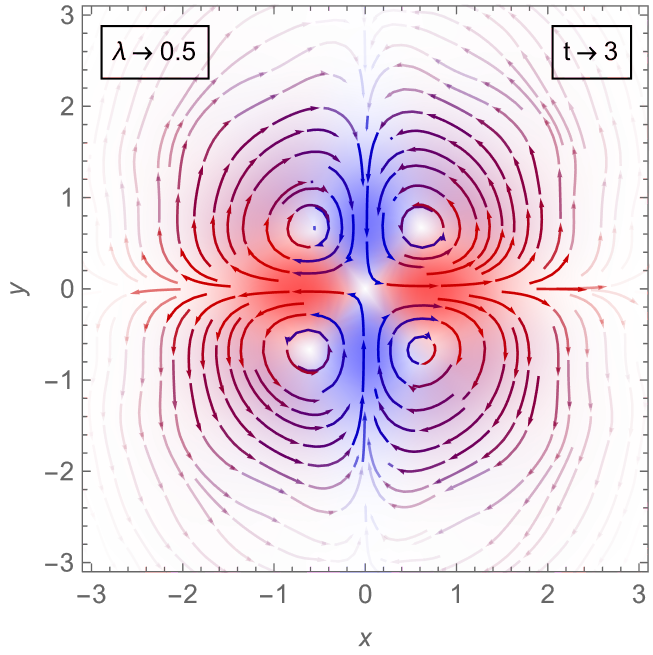}%
\includegraphics[width=.4\columnwidth]{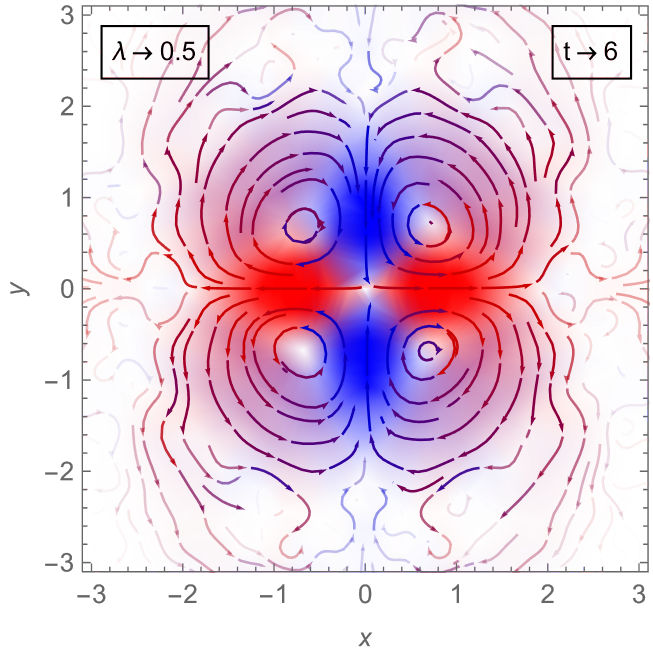}%
}
\vspace*{-2ex}
\caption{Two dimensional probability current $\vec j$, at different times. The current is computed in the truncated unperturbed space regulator for $n=40$, ensuring that the considered times are within the reliability range.  Red and blue colors denote the outgoing and incoming probability flux intensity.}
\label{fig:current}
\end{figure}
%%%%%%%%%%%%%%%%%%%%%%%%%%%

The standard continuity equation  $\dot\rho=\vec\nabla\!\cdot\!\vec j$ holds, for the probability density $\rho(x,y;t)=\psi^*\psi$  and current $\vec j(x,y;t) = \{j_x,j_y\}$, with (in our conventions)
\begin{equation}
j_x = \Im \big(\psi^* \partial_x\psi
\big) \,,\qquad j_y = -\Im \big(\psi^* \partial_y\psi\big) \,,
\end{equation}
where the minus sign in $j_y$ follows from the ghost nature of the $y$ kinetic term. In this form, $\vec j$ describes  the probability flow in the $x$, $y$ space, and should account from the depletion of the  initial would-be ground state.

In Figure~\ref{fig:current},
we depict this 2D flow for small and large $\l$, selecting an intermediate and a large time, $t=3$, 7. These values correspond roughly to the initial decay and the first bounce in the $P(t)$ (see e.g.\  Figure~\ref{fig:LOGP}).
What we find is, after short time, the onset of dominant rotational flows in each quadrant, where the probability  exits along the $x$ axes and enters along $y$.

For small $\l\sim 0.1$, where the survival $P(t)$ shows an oscillating behavior, this flow reverts after half period, with the flow entering along the $x$ axis, and $P(t)$ grows again. 
The overall loss of $P(t)$ can be traced to some probability flowing outside of the central zone, shown in red color in the Figure. Also the external flow to infinity reverses, and in this half period  $P(t)$ is repleted. We have thus the impression  that for small $\l$, although there is a coupling, the oscillatory behavior of the system prevails, preventing any leak to infinity to happen. 

For large $\l\sim 0.5$, where no oscillations are present, we still find this rotational flow. The flow increases and no inversion happens (at least within our reliability time). The net gradual decrease of $P(t)$ can  be traced to the small probability flow to the external region, which is expelled from the rotational region, and which never comes back. The stronger colors, denoting a larger probability flux, is in agreement with the stronger variations of $P(t)$ as noted before (see e.g.,\ Figure~\ref{fig:LOGP}).

As noted above, for some other values of $\l$ (even larger, e.g.\ $\l=1$) the flow inversion happens again, with oscillations of $P(t)$, suggesting a nontrivial overall pattern, as a function of $\l$.

\subsection{Comparison of regulator schemes}

\begin{figure}[b]
 \centerline{
\includegraphics[width=.95\columnwidth]{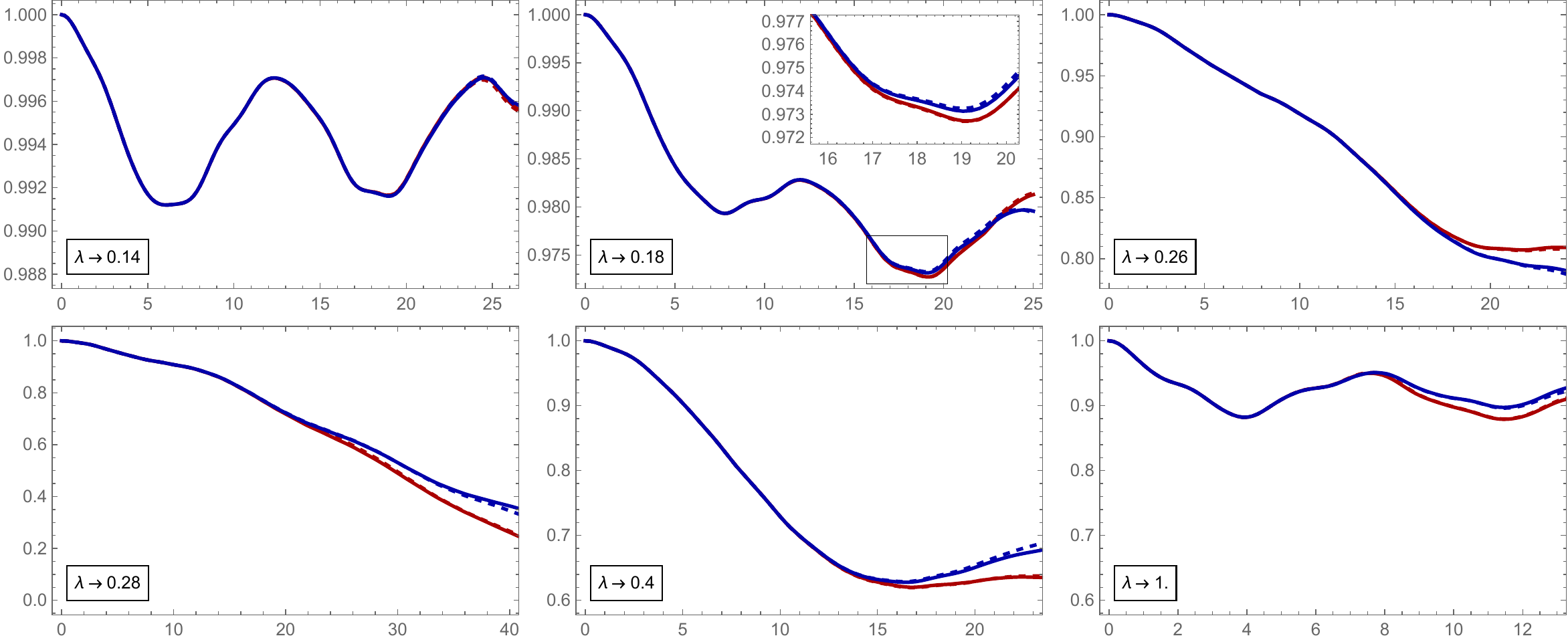}}
\vspace*{-2ex}
\caption{The survival probability as calculated with the two presented regulators. The red and blue lines correspond respectively to the discretized and unperturbed basis schemes. This unveils a quantitative dependence on the regulator scheme, although for small $\l$ the discrepancy is barely visible.} 
\label{fig:comparison}
\end{figure}

Here we compare the two regularization schemes adopted in this work, namely the discretization and the unperturbed truncated basis approaches.
In the previous sections, we have verified that each method  converges numerically when, respectively, the discretization number and the number of states of the truncated  basis (both called $n$) grow.  The separate convergence within each algorithm suggests that the limit for $n\to\infty$ exists, leading to the survival probability in each method. The non sensitivity to the UV cutoff $n$ is welcome and in a sense expected, since the potential is smooth.\footnote{There exist, on the other hand, simple quantum mechanical systems, with contact interactions, which develop a dependence on UV cutoff, and for which a renormalization is necessary before disentangling the IR from the UV degrees of freedom, see for example the lectures~\cite{Luty:2007}.}  

It is then interesting to test whether the two methods converge to the same limit, namely if the limit is independent from the regularization. 
In standard quantum-mechanical problems, one does not expect a dependence on regularization schemes, since the UV details are usually decoupled from the IR dynamics.
In Figure~\ref{fig:comparison} we compare the survival probabilities of the two methods, for a selection of values of $\l$. 
We learn  that, still well within the reliability range of each solution, the results do not agree completely:  the two limits slightly depart from each other.  While in each of the two methods  the results become separately independent from $n$, the direct comparison shows a discrepancy between them.  This discrepancy is larger than finite-$n$ effects within each scheme (dashed lines in the plot).  

From Figure~\ref{fig:comparison}, the discrepancy is less evident for small $\l$, barely visible in the first frame of the Figure, while it becomes macroscopic for large $\l$. We stress nevertheless that even for small $\l$, the discrepancy is present also at small times, even if not visible in the plots: its magnitude is always larger than the finite-$n$ effects within each result.

Thus, we find some scheme dependence for the coupled ghost problem. 
Such result is not unexpected, for instance a sensitivity to UV details of the spectrum is present also in singular potentials in quantum mechanics~\cite{Case:1950an,Meetz62,Perelomov:1970fz}, with the spectrum becoming unbounded from below. Here, the interpretation that we offer is as follows. The ghost system intrinsically mixes IR and UV regimes, and this happens already at the level of the decoupled system. For example, one knows from Section~\ref{sub_ghost_QM_dec} that the spectrum continuously reshuffles when $n$ grows, leading to a dense spectrum for $n\to\infty$. In other words, the UV effects related to large $n$ are genuinely connected to the energy levels at small energies.
This peculiarity induces a sensitivity to the UV details,  resulting in the slight discrepancy between the results in Figure~\ref{fig:comparison}.

In particular, in the discretization method, the lattice step was chosen so that the momentum cutoff and volume scale respectively as  $ a^{-1}\propto n^{1/2}$ and $L\propto n^{ 1/2}$, corresponding to a specific path in the UV-IR cutoffs plane. Specifically, we verified that the results are sensitive to the parameter $C$ in \eqref{cutoff}. 
The method based on the truncated unperturbed basis follows effectively a different path in the UV-IR plane: the volume is intrinsically infinite, $L\to \infty$, whereas the UV regulator $n$ of the truncated energy spectrum corresponds to an effective momentum cutoff scaling again as $\sim n^{1/2}$.  Thus one may interpret this regulator as if we performed the IR limit $L\to\infty$ before the UV one. Evidently, the intrinsic sensitivity to the UV effects translates to direction dependence in the UV-IR cutoffs plane, which induces a slight difference in the physical results. 
In general, one may also imagine other regulator schemes, e.g.\ different cutoffs in the $x$ and $y$ systems, or even  some nontrivial cutoff contours, etc.

The key question becomes: what are the physical consequences of the scheme dependence as uncovered in Figure~\ref{fig:comparison}? In principle, one faces the general problem of matching with physical observations, so let us suppose that one has experimental data for the survival probability, which is the observable under consideration.  In any chosen scheme, one would determine the parameters of the model from some experimental observations: a single observation would suffice to determine $\l$ in the truncated scheme, and two observations would be needed to determine $\l$ and $C$ in the discretization scheme. 
The remaining time dependence of $P(t)$ would then represent testable, theoretical predictions. These would generally differ from one scheme to another, therefore, the regulator effectively becomes part of the model definition rather than merely a computational device. The bottom line is that different regulator choices correspond to different effective models.
Moreover, since one could implement infinitely many possible regulator schemes, one is led to conclude that there is a lack of predictivity regarding all the observables, such as $P(t)$. The latter may depend on an infinite number of independent  parameters---analogous to an effective, nonrenormalizable quantum field theory. In our system, it remains an open problem whether the UV scheme dependence reflects the presence of infinitely many parameters, or a finite set of combinations would effectively suffice to determine the dynamics, thus restoring the predictivity. A pragmatic approach is to simply define the model within some "minimal regulator scheme"---e.g, involving the smallest possible number of parameters---then confront it with phenomenology.

It is worth mentioning that the nontrivial dependency on $\l$ suggests that some specific values of the coupling lead to a peculiar behavior due to resonant or accidental configuration of the diagonalized finite matrix. It is then interesting to analyze also the classical behavior to investigate possible correspondences  with the quantum problem. This is the aim of the following section.

\section{Connection with classical motion?} 
\label{sec:classical}

In this section, we study the classical evolution of the system. In particular, we focus on the large potential, which shows classical solutions with blow up or runaway.\footnote{The small potential instead shows stable orbits: the motion becomes effectively harmonic far from the origin.}

The classical equations of motion, derived from Hamiltonian~\eqref{eq:system} in the large potential case, read
\begin{align}
& x''(t)+x(t)+\lambda\,  x(t)\, y(t)^2=0 \nonumber \\
& y''(t)+2y(t)-\lambda \, x(t)^2\,y(t)=0\,,  \label{classicaleqs}
\end{align}
where the derivation is with respect to time $t$, and where we use the values of $\omega_i$ as in \eqref{free_spectrum_2}. 

The equations possess a rescaling symmetry 
\begin{equation}
\label{eq:rescaling}
x\to k\,x\qquad y\to k\,y\qquad \l\to k^{-2}\,\l\,,
\end{equation}
relating equivalent orbits with different $\l$ and rescaled coordinates.  The scaling  also applies  to the classical energies, in particular the quantities $\l\,E_{1,2}$ are invariant under this rescaling. This symmetry could be exploited to set $\l=1$, but because it is valid only at classical level, we choose to keep $\l$ independent in this section.\footnote{As well-known, extending the symmetry to quantum level requires an additional rescaling of $\hbar$, therefore not recovering equivalence of the quantum evolution---see for example pp 223--225 of~\cite{coleman_1985}.}

We integrate numerically the Equations~(\ref{classicaleqs}),  with the aim of inspecting when the system exhibits a runaway or blow-up.
The stability at the quantum level, suggested  for some small values of $\lambda$ as in Figure~\ref{fig:Prob} or~\ref{fig:LOGP}, raises the natural question of what happens at the classical level for corresponding values of the coupling, even if the energies and the evolution can be compared only qualitatively. 
As initial conditions  for \eqref{classicaleqs} we can choose for instance
\begin{equation}
 x(0)=y(0)=0 \,,\qquad 
 \frac{1}{2} x'(0)^2=\frac{1}{2} y'(0)^2=E_1=E_2 \,,  \label{initial}
\end{equation}
where the first conditions fix the starting point at the origin and the second ones set equal initial energies for the two oscillators (here positive) so that the total energy cancels to zero.  

\begin{figure}[t]
 \centerline{%
\includegraphics[width=.99\columnwidth]{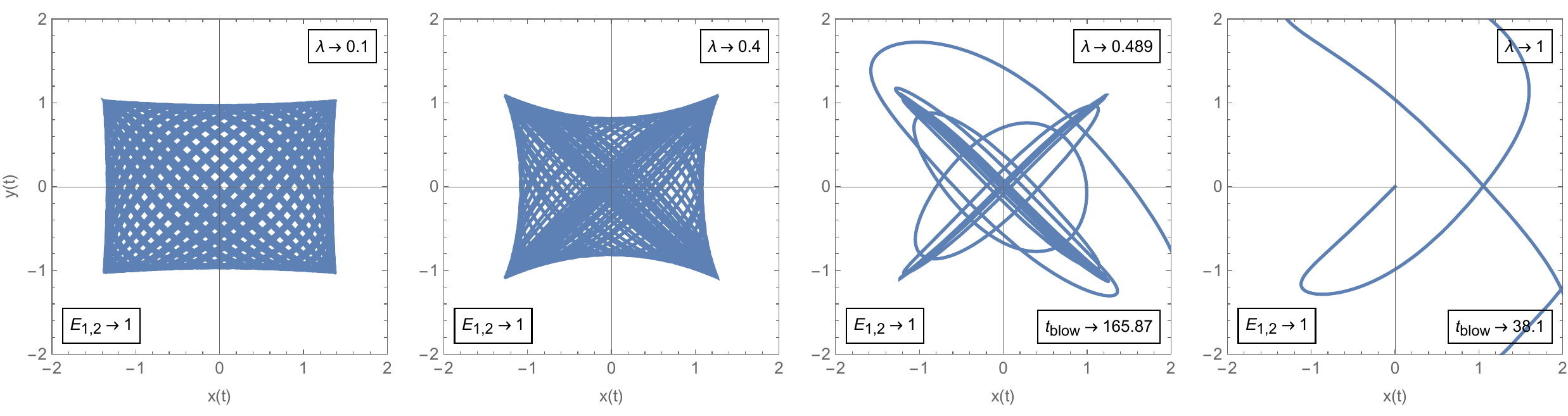}}
\vspace*{-2ex}
\caption{Classical trajectories for  some values of $\l$, and $E_{1,2}=1$ as initial conditions. As $\l$ grows, the motion passes from stable to blow-up behavior,  where we indicate also the time of blow-up.}
\label{Fig_classical1}
\end{figure}

In Figure~\ref{Fig_classical1} we plot some orbits for increasing $\l$. For small $\l$ we see a stable behavior of the orbits, that cover a square in a quasi-periodic motion, up to very large times, beyond the computational limits.  There is thus no apparent onset of instability, or more likely it is exponentially late.   The case of $\l=0.1$ may be compared to the quantum analog, e.g.\ of figure~\ref{fig:Prob} which also shows no decay of the initial state.   Therefore, for small $\l$ and/or small initial energies,  both the quantum and classical systems  display (meta)stability regions, that we will discuss  in what follows. According to the naming in \cite{ErrastiDiez:2024hfq}, our classical system belongs to the category $L_1$.

For larger $\l$ instead the orbits first deform and then end in a blow-up, at finite time; for instance in the two right frames.  The rightmost frame, with $\l=1$, in view of the  rescaling~\eqref{eq:rescaling}, is equivalent to $\l=0.1$ and $E_{1,2}=10$; therefore,  can be compared  with the quantum case of excited initial states of Figure~\ref{Probhigh}, which  has indeed $\l=0.1$ and $E_{1,2}=10$ and shows a fast decay. 

%%%%%%%%%%%%%%%%%%%%%%%%%
%%%%%%%%%%%%%%%%%%%%%%%%%%%

We thus notice a qualitative agreement between the classical and quantum evolution: sufficiently large $\l$ or initial excited  energies  imply a blow up of the classical trajectories and a fast decay of quantum states.

% %%%%%%%%%%%%%%%%%%%%%%%%%
%  \begin{figure}[t]
%  \centerline{
% \includegraphics[width=.32\columnwidth]{figs/T025.pdf}
% \includegraphics[width=.32\columnwidth]{figs/T03.pdf}
% \includegraphics[width=.32\columnwidth]{figs/T048.pdf}
% }
% \vspace*{-2ex}
% \caption{A classical trajectory for $\lambda=0.1$ and $E_{1,2}=10$ for initial conditions, for different integration times: as time grows, the trajectory runs away.}
% \label{Fig_classical1}
% \end{figure}
% %%%%%%%%%%%%%%%%%%%%%%%%%%%
% \clearpage
% \medskip

\medskip

\begin{figure}
    \centerline{% 
    \includegraphics[width=.99\linewidth]{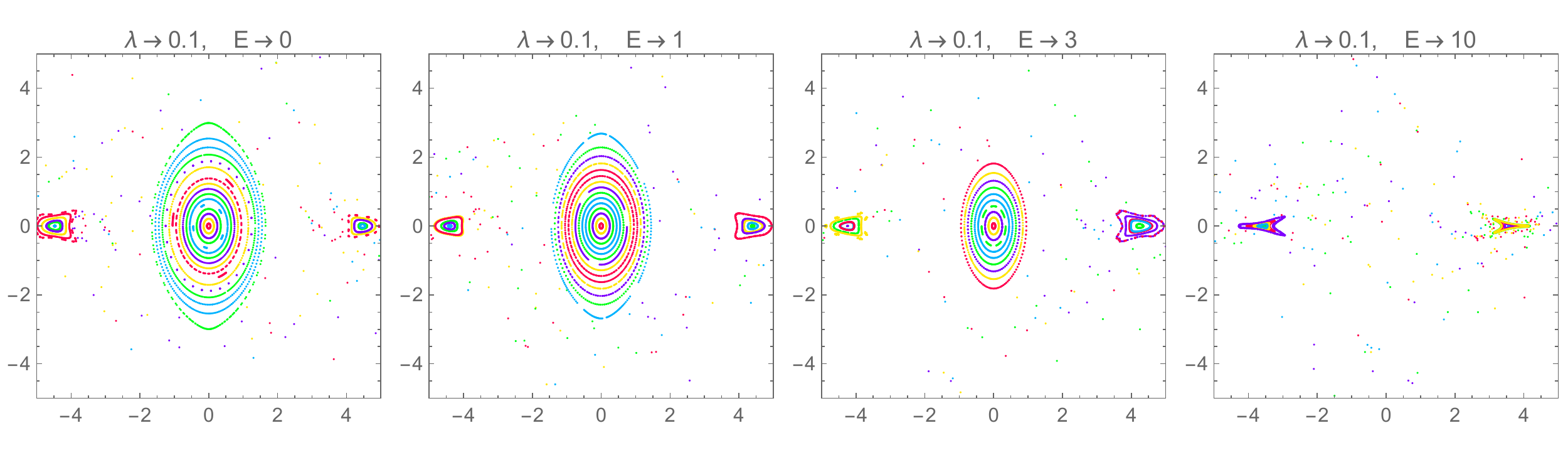}}%
    \vspace*{-2ex}
    \caption{Iterated Poincar\'e maps for $\l=0.1$ and various values for the total energy $E$. In every plot the axes scan the initial conditions $(y(0), y'(0))$, with fixed  $x(0)=0$, and $x'(0)$ determined by the chosen energy. The number of Poincar\'e orbits is 200. The closed contours display the regions of  ``stable'' (quasi) periodic orbits, which are progressively smaller with increasing $E$ (or $\lambda$).} \label{fig:poincaregrid}
\end{figure}

A more precise study of the classical temporal evolution for different initial conditions and for various values of the total energy $E$ (and thus of $\l$ in view of the  scaling symmetry) is provided in Figure~\ref{fig:poincaregrid}, in terms of iterated Poincar\'e maps~\cite{38238}.
Those are defined by choosing first a hyperplane in phase space, here $x=0$; then, fixing a value of the conserved total $E$, for each initial  point on this surface one marks all future intersections of the orbit passing again through  the surface. 
The closed contours then correspond to bounded (quasiperiodic) time evolutions, lasting to large times. Namely, they indicate (quasi)stable orbits, passing from the corresponding initial conditions. 

As we see in Figure~\ref{fig:poincaregrid}, there exists at least one region of (quasi)stability, which is located around the origin, corresponding to the limit of decoupled small harmonic oscillations.   This region shrinks progressively and also disappears for increasing energy $E$. The same happens, again for the scaling~\eqref{eq:rescaling}, for fixed $E$ and increasing $\l$. 

Beyond these (quasi)stability regions, one does not find regions of ergodicity, namely regions filled by aperiodic consecutive Poincar\'e maps. This fact  is understandable because in this system the sections of constant $E$ in phase space are not bounded, due to the ghost nature of $y$.  
In fact, outside the stability regions  the orbits diverge to infinity quite quickly,  namely they blow-up in finite time. 
One finds numerically that this blow-up time shows a fractal pattern in the plane of initial conditions: one has generically very short times, but also filaments of stability passing from one region to the other, etc.  The system therefore shows no ergodicity, but is anyway chaotic. 

% Complementarily, Figure~\ref{fig:sauron} shows the instability time as a function of initial conditions, illustrated for $\lambda=0.1$ and total energy $E=5$. The inner region represents some (quasi)stable Poincar\'e maps, while the external points all lead to unstable orbits. Notice that the instability time shows a fractal pattern.

% \begin{figure}
%     \centering \includegraphics[width=.85\linewidth]{figs/poincaregrid.pdf}
%     \caption{Iterated Poincar\'e maps for various values of $\lambda$ and total energy $E$. In every plot the axes scan the initial conditions $(y(0), y'(0))$, with fixed  $x(0)=0$, and $x'(0)$ determined by the chosen energy. The number of Poincar\'e orbits is 150. The closed contours display the  ``stable'' (quasi) periodic orbits, which are progressively smaller with increasing $E$ or $\lambda$.} \label{fig:poincaregrid}
% \end{figure}

% \begin{figure}
%     \centering \includegraphics[width=.7\linewidth]{figs/sauron.pdf}
%      \vspace*{-1.5ex}
%      \caption{Instability time as a function of initial conditions $(y(0), y'(0))$, for fixed $\lambda=0.1$ and total energy $E=5$. In the inner region we overlay some (quasi)stable Poincar\'e maps. The external points all lead to unstable orbits, with the instability time showing a fractal pattern.}
%     \label{fig:sauron}
% \end{figure}

\begin{figure}
    \centering \includegraphics[width=.75\linewidth]{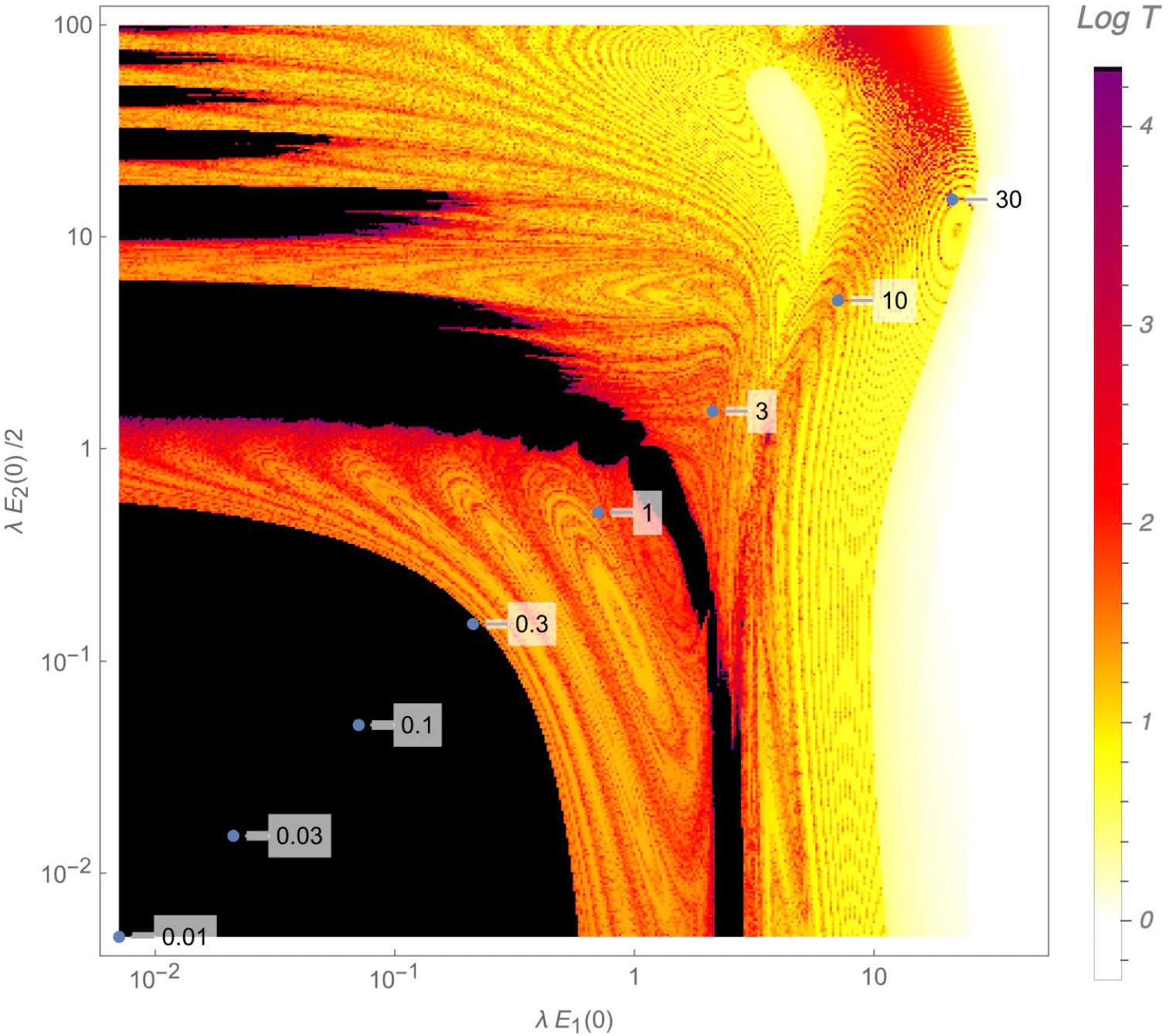}
    \vspace*{-2ex}
    \caption{The color displays the instability time $T$ as a function of rescaled initial energies $\lambda E_1$, $\lambda E_2$. The black regions represent stability times larger than the numerical limit. 
    %The lower-left stability region, for small energies or small $\l$, is well approximated by the condition $\sqrt{\lambda E_1}+\sqrt{\lambda E_2/2}<1$. 
    Entering this region the stability time grows exponentially and overcomes very quickly the numerical time cutoff. 
    On the figure we overlay, for various values of $\lambda$, the points corresponding to the would-be ground state $|0,0\rangle$, having energies $E_{1,2}=\omega_{1,2}/2$. For smaller $\lambda$, they move to exponentially long stability times, supporting the correspondence between quantum and classical quasistability.}
    \label{fig:hole}
\end{figure}

\medskip

We are then interested in studying this stability time for various initial data and $\l$, to compare with the results at quantum level.   
In Figure~\ref{fig:hole} we show  the stability time in terms of the (positive) rescaled energies $\l E_1$ and $\l E_2 /2$ of the initial condition. The plot is thus valid for any value of $\l$.\footnote{We approach this study  by integrating the system in the action/angle variables defined respectively as $E_i/w_i$ and $\Theta_i=\arctan (w_i q_i/p_i)$, which allow for increased numerical reliability.  Figure~\ref{fig:hole} is obtained for $\Theta_{1,2}(0)=\pi/4$; the results for other choices lead to similar results, with slightly increased stability regions.}
A number of results can be drawn from the analysis: %of Figure~\ref{fig:hole}:
\begin{itemize}
    \item for large $\l E_1\gtrsim 30$ the orbits are always unstable, with practically immediate blow-up regardless of $E_2$ (indeed large $x(t)$ has a destabilizing behavior in the second equation of motion~\eqref{classicaleqs}). 
    \item The colored regions show many initial points where the instability time is short, with a definitely nontrivial and fractal pattern.     
    \item There are stability regions, black in the figure, where the stability time exceeds our numerical limits (here $t_{max}=6\times 10^4$). 
    
    When approaching these regions, the instability time grows exponentially very quickly, suggesting either a superexponential increase of the time, or an increase to infinity, namely a region of \emph{real stability}.

    \item Among them, there are stability regions, for small $E_1$ and large values of $E_2$, where due to likely nonresonant situations,  the small $E_1$ is not enough to induce a blow up of the more energetic ghost. (indeed, in the first equation in~\eqref{classicaleqs}, large values of $y(t)$ have a stabilizing detuning effect by accelerating the oscillations of $x(t)$). 

    \item The interesting and expected stability region for small $\l E_{1,2}$ (lower-left in the plot) corresponds to the limit of weak coupling:  for any initial energy, there is a sufficiently small coupling $\l$ so that the evolution is stable. This region can be approximated by the condition $\sqrt{\lambda E_1}+\sqrt{\lambda E_2/2}<1$.
    
\end{itemize}

\medskip

In order to attempt a qualitative comparison of this classical results with the quantum evolution, we recall that for the initial would-be ground state $|0,0\rangle$, the true ground state  initial energies of the two oscillators are $E_{1,2}=\w_{1,2}/2 = \{1/2,1/\sqrt{2}\}$. Therefore, if one takes these values as initial condition for the classical evolution, for different $\l$ they fall in different points in the $\l E_{1,2}$ plane. These are overlaid in Figure~\ref{fig:hole}, and we see that for small  $\l\lesssim 0.3$ they lie inside the stability region, while for larger values they move outside, falling in unstable regions, but passing even in new stable ones.  It is intriguing that a similar pattern is found in the survival probability of the quantum state, e.g.,\ Figure~\ref{fig:LOGP}.

Clearly, one can not push very far this parallel between the classical and quantum evolutions. Nevertheless, one can appreciate here a certain concordance at least between the small $\l$ stability of the classical trajectories and the surviving behavior of the would-be ground state. 

One could guess that this correspondence is justifiable in a path integral approach, because for small $\l$ there is a stability region near the origin of coordinates (as suggested in Figure~\ref{fig:poincaregrid}); so that the very large or infinite classical times needed to exit from this region would prevent the quantum evolution to deplete the wavefunction of the would-be ground state.  For decreasing $\l$ the region enlarges, covering more the initial state wavefunction, in agreement with the result of a larger survival probability, as in~\eqref{eq:estimate}.

%\clearpage

%\section{Discussion}

\section{Summary of the results and discussion}

Let us summarize our findings, focusing on the large potential case since both potentials lead qualitatively to the same conclusions, with the small one featuring even a better numerical convergence.

{%
\renewcommand{\labelenumi}{\Roman{enumi}.}

\begin{enumerate}

\item We find evidence that, when the UV cutoff is removed, also in the coupled system the spectrum remains a dense \emph{point spectrum}, like in the free system. 

While in the free system the spectral density grows linearly with the UV cutoff, in the coupled case we find that the spectral density grows with a lower power of the UV cutoff, a sort of anomalous dimension which depends on the coupling $\l$.

\item The calculation of the survival probability of the initial state requires the overlap coefficients $c_i$.  Although for $\l \gtrsim 0.05$ we do not verify convergence of the $c_i$ when removing the UV cutoff, we \emph{do find} numerical convergence of the physical survival probability $P(t)$.

$P(t)$ shows convergence  within any chosen time interval, $0<t<t_{max}$, with larger values of  $t_{max}$ requiring increasingly larger UV cutoff $n$. For smaller $\l$, one has a faster convergence of $P(t)$, and larger reliability times $t_{max}$.

\item  Despite the dense spectrum and the difficulties discussed, the initial would-be ground state does not decay. Specifically,  at least for some limited values of $\lambda$ where the reliability time is large, the survival probability $P(t)$  remains nonzero at large times with an average magnitude that depends on $\lambda$, and features residual oscillations.

Compared to the metastable well decaying to open infinity, here the outcome resembles more a metastable well decaying into an external region with a further barrier, which prevents the decay to zero and gives rise to residual oscillations. 

The late time behavior of $P(t)$ depends on $\l$: expectedly, the smaller is $\l$, the closer $P(t)$ remains to 1. The asymptotic oscillatory behavior is well verified for small $\l<0.1$, where one can also estimate the residual survival probability,see~\eqref{eq:estimate}.  A similar behavior appears also for larger $\l$, where $P$ tends to deviate significantly from $1$.  However, the pattern  is not regular: for some values of $\l$ (e.g.,\ $\l=0.28$ in Figure~\ref{fig:LOGP}) oscillations are not observed and $P(t)$ decays more but, for some other larger $\l$, the decay reduces again, and residual oscillations reappear, see for example the case of $\l=1$ in Figure~\ref{fig:comparison}.

\item We find that the normal and ghost oscillator exchange energy, but this does not lead to runaway situations: consistently with the oscillations of $P(t)$, there is a finite extraction of energy from the ghost, eventually stabilized in the oscillatory behavior.

\item The analysis of the probability current in the $x$-$y$ plane, Figure~\ref{fig:current}, shows a rotational flux in each quadrant, where initially the flow enters along the (ghost)  $y$ axis and exits along $x$. This flow is reversed in the oscillating phase. The decrease and increase of $P(t)$ is attributed not to this rotational flow, but to the small probability flow in the external region, which also reverses periodically, very similarly to the bounce off a wall. 

When no oscillations are present, the rotational flow is not reversed (within our computable times), and the decay of $P(t)$ in this case is attributed  to the continuous leak of probability in the external regions.

\item We studied what happens when changing $\w_2$, the frequency of the ghost sub-system (keeping $\l$ fixed). As expected, increasing $\w_2$ corresponds to $P(t)$ closer to 1, still with the same qualitative oscillating behavior. This is a sort of ``decoupling'': the harmonic restoration force wins over the destabilizing interaction; namely, the coupling becomes ineffective when the ghost is extremely ``heavy''. 

\item We also analyzed excited initial states. As expected, they show a lower survival probability, as compared to $P(t)$ of the would-be ground state, for the same $\l$.

\item Finally, we uncovered an intrinsic dependence on the regularization scheme used. Even though in each scheme the UV cutoff can be removed leading to a finite unambiguous result, the survival probability $P(t)$ obtained  differs slightly in different regulator schemes.  

We argued that this is a genuine feature of the quantum-mechanical system in consideration, as an in-built sensitivity of the IR to the UV: different schemes are observables because they cut different UV details. One can also realize that in any perturbative series this effect is absent, because at any given perturbative order the UV states are decoupled beyond some cutoff. Therefore, the effect is genuinely non-perturbative. 

\end{enumerate}
}

We add some more comments on these results. 

\medskip

%\paragraph{Semiclassical?} 
%In presenting our strategy to address the problem, we have also discussed its intrinsic difficulties.
%This section aims to explore the interplay between classical, semiclassical, and purely quantum regimes.
A legitimate question, concerns the possibility of connecting to the classical behavior via semiclassical methods. Our understanding of the coupled ghost system is that the answer is negative. 
In fact, the semiclassical  Wentzel-Kramers-Brillouin (WKB) approximation works well in the regime of large $S/\hbar$, with $S$  the classical action. However, here, the cancellation between $H_1$ and $H_2$ holds also for the actions of the two subsystems, implying that the there is no regime where the total action can be considered large.  As is evident in the free system, one can obtain  zero action for arbitrarily large excitation numbers $n_1$, $n_2$. The same happens in the coupled system, due to the negative kinetic term. One can thus conclude that the WKB approximation is of little use,  except perhaps for such small couplings and small time intervals that motion is approximately decoupled.

A second observation is more technical and is related to the unbounded nature of the motion.  A semiclassical approach might rely on Bohr–Sommerfeld quantization, connecting with the orbits of the classical system. However, this method requires (quasi)periodic orbits; for example, for each cycle of the tori in phase space,
\begin{equation}\label{BS}
\oint p(q)\, dq = 2\pi\hbar \left(n + \frac{\nu}{4} \right) \,,
\end{equation}
with $\nu$ the Maslov index, counting the number of turning points.
However the present classical system exhibits open trajectories (runaway or blow-up solutions) making \eqref{BS} inapplicable. In particular, also the methods applied in chaotic systems like the Gutzwiller trace formula \cite{gutzwiller1971periodic} become ineffective. All this is in agreement with the numerical evidence that the spectrum remains smoothly dense in the UV limit, with no appearance of isolated quantized peaks in the spectral density.

%Moreover, the classical trajectories are chaotic and possess nodes, making the notion of semiclassical paths ill-defined—even if closed orbits were present. 

Nevertheless, the fact that one can not identify a formal connection with the classical system, makes even more intriguing the qualitative correspondence that we observed in the previous Section, between regions of numerically stable classical motion and numerically stable would-be ground state, as well as correspondence between instability of more energetic classical orbits and instability of quantum excited states. 

\medskip

%\paragraph{Thermodynamics?} 
Going beyond the problem of a single quantum mechanical ghost studied in this work, the next step would be to consider a possible ensemble of such systems, and investigate the thermal state. In this physically important problem, one faces the typical problem of systems with spectra unbounded from below, namely that a thermal state can not be defined with the Boltzmann weight and a single temperature, because the partition function Tr ${\rm e}^{-\beta H}$ is ill defined if $H$ is unbounded. 
It is nevertheless well known that systems with spectrum which is bounded from below(above), admit positive(negative) temperature. Here, \emph{in the decoupled limit}, the two subsystems have spectra respectively bounded from below (the normal oscillator) and from above (the ghost one). Therefore they can have respectively positive and negative temperatures, $T_1>0$, $T_2<0$. This state is clearly acceptable in the decoupled system. 
These two facts then suggest that even in the presence of a coupling, the only viable situation is obtained by assuming $T_1>0$ different from $T_2<0$, and \emph{no overall thermal equililbrium} among the subsystems.  The vanishing flow of energy between subsystems uncovered in section~\ref{sec:energy} supports the stability of this scenario.

\medskip

A significant, even more nontrivial step would be the transition from quantum mechanics to quantum field theory. 
QFT can be understood as a realization of a quantum mechanical system at every point in space (e.g.\ in a lattice approach) which introduces a substantial increase in complexity. 

On one hand, the IR-UV mixing effects mentioned above are expected to dramatically affect the notion and behavior of the emerging ghost particles with respect to the removal of the UV cutoff; on the other hand, while the standard infinite phase space vacuum decay argument applies to light ghosts (see e.g.~\cite{Cline:2003gs}), heavier such exotic states appear to still satisfy the decoupling limit discussed around Figure~\ref{Figdiffomega}, a fact that offers some reassurance. The QFT case with  $\w_2$ increasingly larger than $\w_1$ corresponds to a heavy ghost field coupled to a lighter standard one. The greater is the ghost mass, the weaker is the interaction with the standard field, thereby possibly (meta)stabilizing the would-be ground (vacuum) state. This scenario may be particularly relevant for models such as Graviweak and GraviGUT of nonrenormalizable gravity, mentioned in the Introduction, where hypothetical ghosts can have masses near the Planck scale, potentially rendering these models physically viable.

\section{Conclusion}

% First, summary
Motivated by the intriguing suggestions that exotic ``ghost" states with negative energies might exist at high scales, we have provided an in-depth analysis on a quantum mechanical model including an oscillator with a negative Hamiltonian coupled to a normal one. We have argued that, although a drastic simplification, this prototype still provides insights into the meaningfulness of ghosts at the quantum level, and implicitly support the possible viability of QFT  models, sympathetic with scenarios such as Graviweak and GraviGUT, or higher derivative renormalizable gravity. 

Let us briefly summarize our methods and findings. We considered a non-positive-definite Hamiltonian in quantum mechanics, composed of two coupled harmonic oscillators, one of them with negative Hamiltonian, and focused on the question whether an initial state would decay due to the coupling between positive and negative energies. We focused in particular on an initial state prepared in the ground state of both oscillators, called the would-be ground state.

We first reviewed the decoupled case, showing that for incommensurable oscillator frequencies the spectrum becomes dense on the whole real axis, when lifting the UV cutoff, with  the spectral density which becomes constant in energy.
This is already a sign of the peculiar impact of UV physics on any energy scale. 

Then, we studied the coupled problem, considering two kinds of nonderivative interaction potentials: one bounded but nonpolynomial; the other, biquadratic, unlimited at infinity.  As expected, we found that the limited potential is easier to handle since it shows a better numerical convergence. The results are, however, similar. 
We attacked the problem with several approaches, studying the perturbation theory and solving fully the systems with two distinct numerical methods, both  involving a numerical UV cutoff. 

% coupled spectrum: UV-IR mix
The results show that even in the coupled case the spectrum continues to increase its density when lifting the UV cutoff, thus maintaining the UV-IR mixing of the free system. 
% convergence but scheme dependence
Despite this peculiarity, the would-be  ground state shows a behavior in time that converges to a definite function, when removing the UV cutoff. However, we uncovered a mild dependence of the limiting curve on the regularization scheme, which can be interpreted as a consequence of the UV-IR mixing---a novel peculiarity, for a quantum-mechanical system.

% non decay -> oscillations
The time evolution of the initial would-be ground state shows an unexpected stability: the general behavior is that the survival probability does not show sign of definite (or exponential) decay, on the contrary it enters in an oscillatory behavior around a constant value. This residual (large time) survival probability  depends on the coupling $\l$ and ranges, for the large(small) potential, from  roughly  $P> 99.9\%$ for $\l<0.1(1.)$ to $P\sim 10\%$ at $\l\sim 1.(100.)$ and to  smaller values beyond. Thus the probability to find the initial state at any later time, $t$, is never zero. 

Our interpretation is as follows. 
Common lore would suggest that an initial state has to decay, since probability can leak into an infinity of states with negative energy, which are all inevitably coupled.  Instead, the oscillatory behavior that we found implies that some probability bounces back into the initial state, and the dispersion of the probability never happens totally. This mechanism was studied by displaying the 2D probability currents. The situation is thus more reminiscent of a metastable decay from an asymmetric double well, in which the initial decay from the higher well eventually settles into oscillations among the two wells. This effect is particularly pronounced for small values of the parameter $\lambda$, but also for certain large values.

We should remark that this protection against decay is a quantitative statement, which we could observe only within the limits of the numerically reliable time ranges. 
Therefore, it remains open the quantitative possibility that at exponentially larger times the situation could  deteriorate and lead to a real decay. 
This possibility would only imply that the would-be ground state is metastable with a high lifetime -- usually a positive feature of an unstable system. 
While this suffices to summarize the central message of the present paper, we conclude by highlighting some other remarkable concept in what follows. 

The system we analyzed lacks closed classical trajectories—or has them only approximately, and in limited phase space regions. In other regions it has open and blow-up behavior. As discussed, this rules out the applicability of the Bohr–Sommerfeld quantization condition. This is not unexpected, as unbounded systems typically exhibit a continuous spectrum.  Nevertheless, our findings point to a reassuring qualitative classical-to-quantum correspondence. In practice, classical runaway parameters are qualitatively associated with very small residual survival probability of the quantum state.
At the same time, one observes a correspondence between the cases of (quasi)stable classical trajectories and (quasi)stable would-be ground state in the  quantum systems. 

\medskip

Briefly, in the ghost system we studied, despite its effectively continuous spectrum, we numerically observe a non-decaying survival probability. This analysis thus supports the possibility that ghost systems may be considered more healthy than previously believed, 
at least at the quantum-mechanical level.

\section*{Acknowledgments}

Work partially supported by the PRIN PE9 project 2022EEZRR3. We acknowledge useful discussions with Jeffrey Rauch and Francesco Di Menna.

\bibliographystyle{unsrt}
\bibliography{biblio}

\end{document}